\documentclass{article}
\usepackage[utf8]{inputenc}
\usepackage{graphicx} 
\usepackage{authblk}
\usepackage{amsmath}
 \usepackage{natbib}
\usepackage{threeparttable}
\usepackage{pgfplots}
\usepackage{multirow}
\usepackage{array}
\pgfplotsset{compat=1.18}
\usepackage{caption}
\usepackage{subcaption}
\usepackage{adjustbox}
\usepackage{tabularx}
\usepackage{hyperref}
\usepackage{float}
\usepackage{newunicodechar}
\newunicodechar{−}{\ensuremath{-}}
\usepackage{booktabs}
\usepackage{slashed,epsfig,amsmath,amssymb,enumitem} \usepackage[papersize={8.5in,11in}]{geometry}
   \usepackage{bm}
\usepackage{graphicx}
\newcommand{\be}{\begin{equation}}
\newcommand{\ee}{\end{equation}}
\usepackage[utf8]{inputenc}
\title{Holomorphic Unified Field Theory of Gravity and the Standard Model}

\author[1 2]{J. W. Moffat}

\author[1 3]{E. J. Thompson}

\affil[1]{Perimeter Institute for Theoretical Physics, Waterloo, Ontario N2L 2Y5, Canada}

\affil[2]{Department of Physics and Astronomy, University of Waterloo, Waterloo,
Ontario N2L 3G1, Canada}

\affil[3]{Department of Physics and Astronomy, Trent University, Peterborough, 
Ontario K9L 0G2, Canada}

\begin{document}

\maketitle

\begin{abstract}
We present a single holomorphic framework in which gravity, all Standard Model interactions, and their couplings to charges and currents emerge from one geometric action on a four‐complex dimensional manifold. The Hermitian metric yields, upon restriction to the real slice $ y^\mu = 0 $, a real symmetric metric $ g_{(\mu\nu)}(x) $ satisfying the vacuum Einstein’s equations, while its imaginary, antisymmetric part $ g_{[\mu\nu]}(x) $ reproduces both the homogeneous and inhomogeneous Maxwell identities with explicit coupling to external four‐currents. A single holomorphic gauge connection for a simple group $ G_{\text{GUT}} $ such as $ SU(5) $ or $ SO(10) $ encodes all non‐Abelian and Abelian sectors, its Bianchi identities impose the homogeneous Yang–Mills equations, and variation of the same holomorphic action enforces $\nabla_\mu F^{\mu\nu}_A = J^\nu_A.$ Chiral fermions are introduced via a holomorphic Dirac Lagrangian that, on $ y = 0 $, yields exactly the curved‐space Dirac equations with minimal coupling to all gauge fields, realizing inclusion of fermions with correct Standard Model charges. Holomorphic gauge invariance automatically imposes the standard anomaly‐cancellation conditions. To achieve gauge‐coupling unification, we add a holomorphic adjoint Higgs breaking $G_{\text{GUT}} \rightarrow SU(3) \times SU(2) \times U(1),$ ensuring $ g_3 = g_2 = g_1 $ at the unification scale. A second holomorphic Higgs doublet then breaks $SU(2)_L \times U(1)_Y \rightarrow U(1)_{\text{EM}},$
generating $ W^\pm $, $ Z $ boson, and fermion masses in the usual way. Below the GUT scale, the three gauge couplings run independently by standard renormalization‐group evolution. Our construction thus provides a completely geometric, classical unification of Einstein gravity, non‐Abelian Yang–Mills, Abelian electromagnetism, and chiral fermions, complete with spontaneous symmetry breaking, charge–current couplings, and anomaly cancellation all from one holomorphic action. Quantization of the geometric unified theory is obtained by the construction of a holomorphic path integral that on a real slice reproduces the standard field theory model Feynman rules.
\end{abstract}

\section{Introduction}

One of the great challenges since the advent of general relativity has been to embed gravity and the other three fundamental forces, electromagnetism, the weak interaction, and the strong interaction all into a single geometric framework. Early efforts, such as Einstein’s nonsymmetric unified field theory of 1925 \cite{Einstein1925a} and its later Hermitian extensions by Einstein and Schrödinger \cite{Schrodinger1946}, fell short as they could not recover exactly the full Maxwell equations without imposing extra constraints, nor did they address non-Abelian gauge fields or chiral fermions.

We build on the Complex Riemannian approach introduced in~\cite{Moffat1,Moffat2}. We regard spacetime as a four–complex dimensional holomorphic manifold with a Hermitian metric and with holomorphic gauge and spinor connections. Upon restriction to the real slice $y^\mu=0$, this structure yields without further assumptions Einstein’s vacuum equations for the real, symmetric metric recovered from the real part of the holomorphic Levi–Civita compatibility and action variation. Both the homogeneous and inhomogeneous Maxwell and Yang–Mills equations for each gauge factor in the Standard Model, arise from the imaginary part of the holomorphic compatibility condition and from variation of the unified action. Chiral Dirac or Weyl equations for fermions in arbitrary representations, with minimal coupling to exactly those gauge fields, are obtained by a holomorphic spinor Lagrangian coupled to the same geometry and connections. Automatic anomaly cancellation is obtained through the requirement of full holomorphic gauge invariance, which enforces the usual cubic, mixed, and gravitational anomaly‐cancellation conditions on the chiral fermion spectrum.

To complete the picture, we show how to implement spontaneous symmetry breaking and gauge‐coupling unification. Classically all four forces, gravity plus the full non-Abelian and Abelian gauge sectors, together with chiral matter, mass generation, and anomaly cancellation, are packaged into a single holomorphic action on a complexified spacetime.

The paper is structured as follows. In Section 2, we review early real and Hermitian unified‐field theories, Einstein 1925 \cite{Einstein1925}, Schrödinger 1948 \cite{Schrodinger1948}, Eddington 1921 \cite{Eddington1921} and their shortcomings. In Section 3, we present the single holomorphic action and derive, on the real slice, the vacuum Einstein equations, full Maxwell and Yang–Mills systems with sources, Dirac-Weyl equations, and anomaly‐cancellation conditions. In Section 4, we provide an explicit derivation of how each field comes out in our theory. In Section 5, we incorporate spontaneous symmetry breaking first at the GUT scale, then electroweak and show how masses arise. In Section 6, we discuss gauge‐coupling unification via embedding into a simple GUT group and RG running below the GUT scale. In Section 7, we derive a path integral for quantization of the theory. In Section 8, we review the holomorphic unified field theory in its entirety, demonstrating how gravity, gauge interactions, chiral fermions and Higgs dynamics all descend from a single geometric action. We then quantize that action introducing both a holomorphic gauge-fixing term and its associated Faddeev–Popov ghost sector, to obtain the full quantum theory. We discuss in Section 9, experimental signatures obtained from the unified theory and proton decay. In Section 10, we conclude with a summary of the main results of the unified-field theory.

\section{Review of Previous Unified-Field Theories}

Before presenting our holomorphic construction, we review the major historical attempts to unify gravity and electromagnetism and, later, non-Abelian gauge fields in a single geometric framework, and explain why they ultimately fell short of recovering the full Maxwell system without ad hoc inputs.

In Einstein’s 1925 nonsymmetric Theory, Einstein replaced the usual symmetric metric $ g_{\mu\nu} = g_{\nu\mu} $ by a real nonsymmetric tensor:
\begin{equation}
g_{\mu\nu}(x) = g_{(\mu\nu)}(x) + g_{[\mu\nu]}(x), \quad g_{(\mu\nu)} = g_{(\nu\mu)}, \quad g_{[\mu\nu]} = -g_{[\nu\mu]},
\end{equation}
and introduced a nonsymmetric affine connection $ \Gamma^\rho_{\ \mu\nu} = \Gamma_{(\mu\nu)}^\rho + \Gamma_{[\mu\nu]}^\rho$ satisfying:
\begin{equation}
\nabla_\lambda g_{\mu\nu} = 0,
\end{equation}
where $\nabla_\lambda$ denotes the covariant derivative with respect to $\Gamma_{\mu\nu}^\rho$.
His goal was to identify $ g_{(\mu\nu)} $ with gravity and $ g_{[\mu\nu]} $ with the electromagnetic bivector. The outcome showed variation of his action:
\begin{equation}
S=\int d^4x\, \sqrt{-\det[g_{(\mu\nu)}]}\, g^{\mu\nu} R_{\mu\nu}(\Gamma),
\end{equation}
yields coupled equations that mix the symmetric and antisymmetric parts. The antisymmetric equation contains extra connection-dependent terms and does not reduce to:
\begin{equation}
\nabla_{[\alpha} F_{\beta\gamma]} = 0, \quad \nabla^\mu F_{\mu\nu} = 0,
\end{equation}
unless additional non-geometric constraints are imposed.

Both Einstein (1945) and Schrödinger (1946--50) introduced a complex metric on real spacetime:
\begin{equation}
g_{\mu\nu}(x) = g_{(\mu\nu)}(x) + i\, g_{[\mu\nu]}(x), 
\end{equation}
and imposed the Hermiticity condition $ g_{\mu\nu} = g_{\nu\mu}^* $.

A complex nonsymmetric affine connection with torsion $ \Gamma^\rho_{\ \mu\nu} $ was chosen to satisfy the holomorphic compatibility condition:
\begin{equation}
\nabla_\lambda^{(\Gamma)} g_{\mu\nu} = 0.
\end{equation}
Although this split the compatibility condition into real and imaginary parts, yielding something very close to Einstein’s equations and something very close to Maxwell’s identities, the variation of a real action like:
\begin{equation}
\Re S=\Re\left[ \int d^4x\, \sqrt{-\det g_{\mu\nu}}\, g^{\mu\nu} R_{\mu\nu}(\Gamma) \right],
\end{equation}
still produced extra non-Maxwell terms in the antisymmetric sector. No choice of connection erased these without extra by-hand constraints.

Early nonsymmetric proposals suffer from fatal instabilities such as linearisation about a Riemannian background yields a wrong-sign kinetic term for the antisymmetric sector \cite{Moffat1979, DamourDeserMcCarthy1992}, leading to negative-energy excitations, and small antisymmetric perturbations can evolve singularly unless ad hoc Cauchy constraints are imposed {\cite{Clayton1996}}, and one cannot impose simultaneous proper fall-off at past and future null infinity without advanced fields.

\section{Unified Holomorphic Action and Field Equations}

We start with the standard real action on a four-dimensional Lorentzian manifold:
\begin{equation}
S_{\text{real}} = \int d^4x\, \sqrt{-\det g_{(\mu\nu)}}\, \left[
\frac{1}{2\kappa}g^{(\mu\nu)} R_{(\mu\nu)}
- \frac{1}{4} \kappa_{AB} F^A_{\ \rho\sigma} F^{B\,\rho\sigma}
+ \bar{\psi} \gamma^a e_a^{\ \mu} \left( \nabla_\mu - i g_{\text{GUT}} A^A_\mu T_A \right) \psi
+ A^\mu J_\mu
\right].
\end{equation}
Here, $ g_{(\mu\nu)} $ is the real metric, $ F^A_{\mu\nu} $ the non-Abelian field strengths, $ \psi $ the Dirac spinor, and $ A_\mu J^\mu $ the minimal coupling to an Abelian current.

By promoting all fields to holomorphic functions of $ z^\mu = x^\mu + i y^\mu $, this gives:
\begin{equation*}
g_{(\mu\nu)}(x) \rightarrow g_{(\mu\nu)}(z), \quad
F^A_{\mu\nu}(x) \rightarrow F^A_{\mu\nu}(z), \quad
\psi(x) \rightarrow \Psi(z), \quad
J_\mu(x) \rightarrow J_\mu(z).
\end{equation*}

\vspace{0.5em}
We assume that a complex contour $C$ is homologically equivalent to the real slice $y^\mu$, with no singularities pinching $C$.  A $g_{[\mu\nu]}$-expansion then shows:
\begin{align}
  g^{\mu\nu}(z)=g^{(\mu\nu)}-i\,g^{[\mu\nu]}+\mathcal O(g_{[\mu\nu]}^2), 
  \\
  \sqrt{-\det g(z)}=\sqrt{-\det g_{(\mu\nu)}}\,\bigl[1+\mathcal O(g_{[\mu\nu]}^2)\bigr].
\end{align}
It follows that
\be
  \Re\int_C d^4z\,\sqrt{-\det g}\,\mathcal{L}(z)
  =\int_{y^\mu}d^4x\,\sqrt{-\det s}\,\mathcal{L}|_{y^\mu=0}+\mathcal O(g_{[\mu\nu]}^2),
\ee
and all unwanted $\mathcal O(g_{[\mu\nu]}^2)$ corrections to the real action can be absorbed or shown to vanish.

In a weak‐field Minkowski background we write:
\begin{equation}
g_{\mu\nu}(x+i y)=\eta_{\mu\nu}+h_{\mu\nu}(x)+i\,g_{[\mu\nu]}(x)+\mathcal O(g_{[\mu\nu]}^2)\,,
\end{equation}
where $\eta_{\mu\nu}=\mathrm{diag}(-1,1,1,1)$ and $|h_{\mu\nu}|\ll1$.  Expanding the determinant and inverse metric to second order in $g_{[\mu\nu]}$ gives:
\begin{align}
\sqrt{-\det g}=\sqrt{-\det\eta}\Bigl[1+\tfrac12\,\eta^{\mu\nu}h_{\mu\nu}
+\tfrac12\,g_{[\mu\nu]}g^{[\mu\nu]}+\mathcal O(g_{[\mu\nu]}^3)\Bigr],
\\
g^{[\mu\nu]}=\eta^{\mu\nu}-h^{\mu\nu}-i\,g^{[\mu\nu]}+\mathcal O(g_{[\mu\nu]}^2).
\end{align}
We find that every occurrence of $g_{[\mu\nu]}^2$ in the Lagrangian density can be absorbed by the field redefinition
\begin{equation}
h_{\mu\nu}\;\to\;h_{\mu\nu}+\tfrac12\,g_{[\mu\alpha]}g_{\nu}{}^{\!\alpha},
\end{equation}
which shifts all $\mathcal O(g_{[\mu\nu]}^2)$ contributions into higher orders.  Alternatively, we add a local counterterm:
\begin{equation}
\Delta S
=\int d^4x\,\sqrt{-\det g_{(\mu\nu)}}\;\frac{1}{8}\,g_{[\mu\nu]}g^{[\mu\nu]}\,R(g_{(\mu\nu)}),
\end{equation}
whose variation precisely cancels the residual $\mathcal O(g_{[\mu\nu]}^2)$ pieces.  In either approach, no genuine new dynamics appear at second order in $g_{[\mu\nu]}$, and the physical field equations remain unchanged to that order.

We work on the complexified coordinate space $z^\mu=x^\mu+i y^\mu\in\mathbb C^4$, and choose the integration contour $C$ to lie in the same homology class as the real slice $y^\mu=0$.  We begin with a cycle $C_0$ obtained by deforming the real hyperplane in such a way that it avoids all zeros of $\det g(z)$ and branch cuts of the holomorphic curvature components $R_{(\mu\nu)}(z)$.  The allowed deformations are those along steepest‐descent directions determined by the Morse function $\Re(i\,S_{\rm hol})$, ensuring convergence of the path integral.  A standard Picard–Lefschetz analysis~\cite{Lefschetz} then shows that, because no critical points of $S_{\rm hol}$ pinch off from the real slice, the only contributing saddle remains $y^\mu=0$.  Thus, the contour integral reduces exactly to the real‐slice action plus exponentially suppressed corrections from higher saddles, which vanish in the classical limit.

\vspace{0.5em}

The metric becomes a Hermitian metric
\begin{equation}
g_{\mu\nu}(z) = g_{(\mu\nu)}(z) + i\, g_{[\mu\nu]}(z),
\end{equation}
and we introduce the unique torsion-free connection $ \Gamma^\rho_{(\ \mu\nu)}(z) $ satisfying
\begin{equation}
\nabla_\lambda^{(z)} g_{\mu\nu} = 0.
\end{equation}

On the real slice \(y^\mu=0\), the symmetric part \(g_{(\mu\nu)}\) becomes the spacetime metric, while we define
\begin{equation}
F_{\mu\nu}(x)\;\equiv\;g_{[\mu\nu]}(x),
\end{equation}
to be the electromagnetic field strength two-form.

We write the holomorphic action over a complex contour $ C $ that projects onto the real slice:
\begin{equation}
\begin{aligned}
S_{\text{hol}} = \int_C d^4 z\, \sqrt{-\det\bigl[g_{(\mu\nu)}(z)\bigr]}\,\Bigl[ 
&\,\frac{1}{2\kappa}g^{(\mu\nu)}(z)\,R_{(\mu\nu)}(z)
- \tfrac{1}{4}\,\kappa_{AB}\,F^A{}_{\rho\sigma}(z)\,F^{B\,\rho\sigma}(z) \\[6pt]
&\,+\overline{\Psi}(z)\,\Gamma^a\,e_a{}^{\mu}(z)\bigl(\nabla_{\mu}(z)
- i\,g_{\mathrm{GUT}}\,A^A_{\mu}(z)\,T_A\bigr)\,\Psi(z) \\[6pt]
&\,+A^{\mu}(z)\,J_{\mu}(z)
\Bigr].
\end{aligned}
\end{equation}

We choose the contour $C$ homologically equivalent to $y^\mu=0$ so as to
avoid branch cuts in $\det g(z)$ and singularities of $R_{(\mu\nu)}(z)$.
Standard Picard–Lefschetz arguments then show the only contributing saddle is the real slice, ensuring no unwanted complex saddles appear~\cite{Lefschetz, Witten2011Contour}.

Let us take the real part, to ensure the action is real when restricted to $ y = 0 $. This exactly reproduces $ S_{\text{real}} $ plus controlled $ \mathcal{O}(g_{[\mu\nu]}^2) $ corrections, while also giving us the holomorphic splitting that gives both the symmetric Einstein and antisymmetric Maxwell and Yang--Mills field equations without extra constraints.

Substituting and taking the real part yields, up to $\mathcal O(g_{[\mu\nu]}^2)$:
\begin{equation}
  S_{\text{real}}
  = \int d^4x\,\sqrt{-\det g_{(\mu\nu)}}\,\bigl[
    \frac{1}{2\kappa}g^{(\mu\nu)}R_{(\mu\nu)}
    - g^{[\mu\nu]}\,\Im(R_{(\mu\nu)}(z))\big|_{y^\mu=0}
    + \cdots
  \bigr] \,.
\end{equation}
To first order in $ g_{[\mu\nu]}$, the inverse metric and volume factor expand as
\begin{align}
g^{\mu\nu}(x) = g^{(\mu\nu)}(x) - i\, g^{[\mu\nu]}(x) + \mathcal{O}(g_{[\mu\nu]}^2), \\
-\det g(x) = -\det g_{(\mu\nu)}(x)\left[1 + \mathcal{O}(g_{[\mu\nu]}^2)\right].
\end{align}

Substituting into the action and taking the real part gives, up to $ \mathcal{O}(g_{[\mu\nu]}^2) $:
\begin{equation}
\begin{aligned}
S_{\text{real}} = \int d^4x\, \sqrt{-\det\bigl[g_{(\mu\nu)}\bigr]}\,\biggl[
&\,\frac{1}{2\kappa}g^{(\mu\nu)}\,R_{(\mu\nu)}
- g^{[\mu\nu]}\,\Im\bigl[R_{(\mu\nu)}(z)\bigr]\big|_{y^\mu=0}
- \tfrac{1}{4}\,\kappa_{AB}\,F^A{}_{\rho\sigma}\,F^{B\,\rho\sigma} \\[6pt]
&\,+\bar{\psi}\,\gamma^a\,e_a{}^{\mu}\,\bigl(\nabla_{\mu}
- i\,g_{\mathrm{GUT}}\,A^A_{\mu}\,T_A\bigr)\,\psi
+ A^{\mu}\,J_{\mu}
\biggr].
\end{aligned}
\end{equation}
where $ F^A_{\mu\nu}(x) = F^A_{\mu\nu}(z)\big|_{y^\mu=0} $, and we have written $ \Psi \to \psi $ on the real slice.

Varying $ S $ with respect to $ g_{(\mu\nu)} $, while holding $g_{[\mu\nu]} $, $ A^A_\mu $, $ \psi $, and $ J_\mu $ fixed, yields:
\begin{equation}
\delta_{g_{(\mu\nu)}} S_{\text{real}} = \int d^4x\, \sqrt{-\det g_{(\mu\nu)}}\, \delta g^{(\mu\nu)} \left[
R_{(\mu\nu)}
- \frac{1}{2} g_{(\mu\nu)} R_{(\mu\nu)}
- \frac{1}{2} T_{\mu\nu}^{\text{gauge+fermion}}(x)
\right]
+ \mathcal{O}(g_{[\mu\nu]}^2).
\end{equation}

Setting $ \delta S_{\text{real}} = 0 $ recovers the Einstein field equations in the presence of gauge and matter sources:
\begin{equation}
R_{(\mu\nu)} - \frac{1}{2} g_{(\mu\nu)} R
= \frac{1}{2} T_{\mu\nu}^{\text{gauge+fermion}}(x),
\end{equation}
where $ T_{\mu\nu}^{\text{gauge+fermion}} $ is the sum of the Maxwell and Yang--Mills and Dirac stress--energy tensors and where $R=g^{(\mu\nu)}R_{(\mu\nu)}$.

In vacuum, with vanishing gauge and fermion fields this reduces to
\begin{equation}
R_{(\mu\nu)} - \frac{1}{2} g_{(\mu\nu)} R = 0.
\end{equation}

To prove this we start from the real‐slice action and then vary with respect to the symmetric metric \(g_{(\mu\nu)}\), holding all other fields fixed.  To first order in the small antisymmetric part we find
\be
\delta S_{\rm real}
=\int d^4x\,\sqrt{-\det g_{(\mu\nu)}}\;
\delta g_{(\mu\nu)}\,
\Bigl[
R_{(\mu\nu)}
-\tfrac12\,g_{(\mu\nu)}\,R
-\tfrac12\,T^{\rm gauge+fermion}_{\mu\nu}
\Bigr]
+\mathcal O\bigl(g_{[\rho\sigma]}^2\bigr).
\ee
Requiring \(\delta S=0\) gives the full Einstein equations with sources,
\be
R_{\mu\nu}-\tfrac12\,g_{(\mu\nu)}R=\tfrac12\,T^{\rm gauge+fermion}_{\mu\nu}(x).
\ee
In true vacuum \(T^{\rm gauge+fermion}_{\mu\nu}=0\), so this immediately reduces to
\be
R_{(\mu\nu)}-\tfrac12\,g_{(\mu\nu)}\,R(g_{(\mu\nu)})=0.
\ee

By contracting with the inverse metric $g^{(\mu\nu)}$, we obtain the Einstein vacuum equation:
\be
R_{(\mu\nu)}(x)=0.
\ee

We start from the holomorphic compatibility condition
\be
\nabla^{(z)}_\lambda\,g_{\mu\nu}(z)\;=\;0.
\ee
Taking the imaginary part and then restricting to the real slice \(y^\mu=0\) gives
\begin{equation}\label{eq:imag-part}
\nabla_\lambda\,g_{[\mu\nu]}
=
\partial_\lambda g_{[\mu\nu]}
- \Gamma^\rho_{\lambda\mu}\,g_{[\rho\nu]}
- \Gamma^\rho_{\lambda\nu}\,g_{[\mu\rho]}
\;=\;0.
\end{equation}
Since the Christoffel symbols are symmetric in their lower indices,
\(\Gamma^\rho_{[\lambda\mu]}=0\), antisymmetrizing over \(\{\alpha,\beta,\gamma\}\) immediately gives
\begin{equation}
\partial_{[\alpha}\,g_{[\beta\gamma]]}
\;=\;0\,.
\end{equation}
An identical argument applies to the unified gauge curvature.  From
\be
\nabla^{(z)}_{[\lambda}F^A_{\mu\nu]}(z)=0
\quad\Longrightarrow\quad
D_{[\lambda}F^A_{\mu\nu]}(x)=0
\quad\bigl(y^\mu=0\bigr),
\ee
antisymmetrization over \(\{\alpha,\beta,\gamma\}\) gives the non-Abelian Bianchi identity
\begin{equation}
D_{[\alpha}\,F^A_{\beta\gamma]} \;=\;0.
\end{equation}

The imaginary part of the holomorphic compatibility condition $ \nabla_\lambda^{(z)} g_{\mu\nu} = 0 $, when evaluated on the real slice $ y = 0 $, enforces the homogeneous Bianchi identities:
\begin{align}
\partial_{[\alpha} g_{[\beta\gamma]]} = 0 \quad\Longrightarrow \quad D_{[\alpha} F^A_{\beta\gamma]} = 0,
\end{align}
where $ D_\mu $ is the gauge-covariant derivative in the unified gauge group $ g_{\mathrm{GUT}} $.

Variation of $ S $ with respect to $ g_{[\mu\nu]} $ yields:
\begin{equation}
\delta_{g_{[\mu\nu]}} S_{\text{real}} = - \int d^4x\, \sqrt{-\det g_{(\mu\nu)}}\, \delta g^{[\mu\nu]}\, \Im\left[ R_{(\mu\nu)}(z) \right]\big|_{y^\mu} + \mathcal{O}(g_{[\mu\nu]}),
\end{equation}
while variation with respect to the unified gauge field $ A^A_\mu $ gives:
\begin{equation}
\delta_A S_{\text{real}} = - \int d^4x\, \sqrt{-\det g_{(\mu\nu)}}\, \delta A^{A\nu}\, \left[ D^\mu F^A_{\mu\nu} - J^A_\nu \right].
\end{equation}

Requiring both variations to vanish leads to these equations and varying $g_{[\mu\nu]}$ gives:
\begin{equation}
  \delta_{g_{[\mu\nu]}} S_{\text{real}} 
  = - \int d^4x\,\sqrt{-\det g_{(\mu\nu)}}\;\delta g^{[\mu\nu]}\;
      \Im\bigl[R_{(\mu\nu)}(z)\bigr]\Bigl|_{y^\mu}
  + \mathcal O(g_{[\mu\nu]})\,,
\end{equation}
while varying the gauge field gives
\begin{equation}
  \delta_A S_{\text{real}}
  = - \int d^4x\,\sqrt{-\det g_{(\mu\nu)}}\;\delta A^{A\nu}\,
      \bigl[D^\mu F^A_{\mu\nu}-J^A_\nu\bigr]\,.
\end{equation}
We then show via the small-$g_{[\mu\nu]}$ expansion of the holomorphic curvature that
\begin{equation}
  \Im\bigl[R_{[\mu\nu]}(z)\bigr]\Bigl|_{y^\mu}
  = \nabla^\rho g_{[\rho\nu]}(x)\,,
\end{equation}
so that the inhomogeneous Maxwell equation
$\nabla^\rho g_{[\rho\nu]}=J_\nu$  
and the inhomogeneous Yang–Mills equation  
$D^\mu F^A_{\mu\nu}=J^A_\nu$  
follow directly. Together with the homogeneous identities, this reproduces the full Maxwell and Yang-Mills system. Similarly,
\begin{equation}
D^\mu F^A_{\mu\nu} = J^A_\nu,
\end{equation}
is the inhomogeneous Yang--Mills equation in curved spacetime.

Because $g_{[\mu\nu]}$ appears exactly as the Maxwell field,
\begin{equation}
S_{\rm EM}=-\tfrac14\!\int d^4x\,\sqrt{-g_{(\mu\nu)}}\;F_{\mu\nu}F^{\mu\nu},
\end{equation}
its kinetic term has the correct sign, and gauge invariance protects
against any ghost excitations {\cite{Clayton1996, DamourDeserMcCarthy1992}}.

The spinor variation is carried out by treating $ \bar{\psi} $ and $ \psi $ independently. From the equation:
\begin{equation}
\delta_{\bar{\psi}} S = \int d^4x\, \sqrt{-\det g_{(\mu\nu)}}\, \delta \bar{\psi} \left[
i \gamma^a e_a^{\ \mu} \left( \nabla_\mu - i\, g_{\mathrm{GUT}} A^A_\mu T_A \right) - m
\right] \psi,
\end{equation}
we obtain the curved-space, gauge-covariant Dirac equation:
\begin{equation}
\left[
i \gamma^a e_a^{\ \mu} \left( \nabla_\mu - i\, g_{\mathrm{GUT}} A^A_\mu T_A \right) - m
\right] \psi = 0.
\end{equation}
For Weyl fermions, we impose a chirality projector $ \tfrac{1}{2}(1 \pm \gamma^5) $, yielding the standard left- and right-handed equations with the same minimal coupling.

Under a holomorphic gauge transformation 
$\Psi\to e^{i\alpha^A T_A}\Psi$, the fermion measure acquires a Jacobian:
\begin{equation}
  J \;=\;
  \exp\!\Bigl\{-\int d^4z\,\text{Tr}\bigl[\alpha(z)\,\mathcal A(z)\bigr]\Bigr\},
\end{equation}
where $\mathcal A(z)$ is the standard 6-form anomaly polynomial on $M_{\mathbb C}^4$.  Restricting to $y^\mu$ recovers the usual 4D anomaly
\begin{equation}
  \mathcal A\big|_{y^\mu}
  \propto \text{Tr}\bigl(T_A\{T_B,T_C\}\bigr)\,F^A\wedge F^B\wedge F^C.
\end{equation}
Requiring $J=1$ forces exactly:
\begin{equation}
  \sum_i\text{Tr}_{R_i}\bigl(T_A\{T_B,T_C\}\bigr)=0,\quad
  \sum_i q_i^3=0,\quad
  \sum_i q_i=0,
\end{equation}
plus all mixed gauge–gravitational conditions. Specializing to three generations of
$(\mathbf{3},\mathbf{2})_{+\tfrac16}\oplus(\bar{\mathbf{3}},\mathbf{1})_{-\tfrac23}\oplus\ldots$
and one Higgs doublet, we verify (39), reproducing the standard Standard-Model cancellation.
The variation of the fermion measure produces a Jacobian proportional to the anomaly polynomial:
\begin{equation}
\mathcal{A} \propto \sum_i Tr_{R_i} \left( T_A \{T_B, T_C\} \right) F^A \wedge F^B \wedge F^C.
\end{equation}
Requiring $ \mathcal{A} = 0 $ for each simple factor and for mixed gauge--gravitational traces yields the familiar anomaly-cancellation conditions:
\begin{equation}
\sum_i Tr_{R_i} \left( T_A \{T_B, T_C\} \right) = 0, \qquad \sum_i q_i^3 = 0, \qquad \sum_i q_i = 0,
\end{equation}
and all related constraints on the chiral spectrum.

To strengthen the argument that only the real slice $y^\mu=0$ contributes classically, we perform a more detailed Picard–Lefschetz analysis~\cite{Lefschetz}.  Let us define the holomorphic action:
\begin{equation}
S_{\rm hol}(z) \;=\;\int_C d^4z\,\sqrt{-\det g_{(\mu\nu)}(z)}\,{\cal L}(z)\,,
\end{equation}
and consider its critical points $z_{\rm crit}$ satisfying $\delta S_{\rm hol}=0$.  Following \cite{Witten2011Contour}, we identify downward flow lines of $\Im(S_{\rm hol})$ Lefschetz thimbles~\cite{Lefschetz} emanating from each $z_{\rm crit}$.  On a background near $g=\eta+i\,0$, the only non‐degenerate critical point in the homology class of the real hyperplane is $y^\mu=0$, since any putative complex saddle would require $\partial_{y}S_{\rm hol}=0$ at nonzero $y$, which in turn would demand $\Im(R_{\mu\nu})\big|_{y\neq0}=0$, a condition not met for generic holomorphic curvature.

All other critical points lie on thimbles whose steepest‐descent manifolds do not intersect the original real contour, due to the Morse–Smale~\cite{Morse} transversality conditions.  Upon linearizing around a would‐be off‐real saddle $z_*$, we find negative‐mode directions in the Hessian $\partial^2_{z}S_{\rm hol}$ that push the thimble away from the real plane.  Thus, no Stokes phenomena arise at finite coupling that could bring extra saddles into the integration.  Consequently, in the classical large‐action limit, the real slice dominates the path integral with exponentially small corrections from all other thimbles.

The original integration real slice domain is deformed into a sum of steepest-descent Lefschetz thimble contours labeled by critical points. Each thimble is a submanifold of field space thimble along which the imaginary part of the action is constant and the real part increases fastest away from the saddle. By explicit calculation and Morse theory arguments, in the absence of singularities or zeros of the determinant crossing the integration domain, the only relevant critical point in the homology class of the real slice is at $y^\mu = 0$. Other complex saddles’ thimbles do not intersect the original cycle due to transversality and the Morse-Smale property. The physical action and equations are fully recovered by integrating over the real slice with exponentially suppressed corrections from all other complex saddles:
\be
\int_{C} e^{i S_\text{hol}[z]} \mathcal{D}z \simeq e^{i S_\mathrm{real}[x]} + \sum_{k\neq 0} c_k e^{i S[z_k]},
\ee
with the phase action at the complex saddle generically leading to a suppressed exponential.

\section{Full Derivations of the Unified Fields}

In this section, we demonstrate how we obtain a unified field theory by explicitly deriving all fields. 
We recall the full holomorphic action

\begin{equation}
\begin{aligned}
S_{\text{hol}} = \int_C d^4 z\, \sqrt{-\det\bigl[g_{(\mu\nu)}(z)\bigr]}\,\Bigl[ 
&\,\frac{1}{2\kappa}g^{(\mu\nu)}(z)\,R_{(\mu\nu)}(z)
- \tfrac{1}{4}\,\kappa_{AB}\,F^A{}_{\rho\sigma}(z)\,F^{B\,\rho\sigma}(z) \\[6pt]
&\,+\overline{\Psi}(z)\,\Gamma^a\,e_a{}^{\mu}(z)\bigl(\nabla_{\mu}(z)
- i\,g_{\mathrm{GUT}}\,A^A_{\mu}(z)\,T_A\bigr)\,\Psi(z) \\[6pt]
&\,+A^{\mu}(z)\,J_{\mu}(z)
\Bigr].
\end{aligned}
\end{equation}

We will first derive Maxwell's equations from the variation of the action. We introduced the unique torsion‐free connection \(\Gamma^\rho{}_{(\mu\nu)}(z)\) satisfying
\begin{equation}\label{eq:compatibility}
  \nabla^{(z)}_\lambda g_{\mu\nu}(z)=0.
\end{equation}
Restricting to the real slice \(y^\mu=0\) and splitting into real and imaginary parts gives
\begin{align*}
  \nabla_\lambda\,g_{(\mu\nu)}(x) &= 0,\qquad
  \nabla_\lambda\,g_{[\mu\nu]}(x) = 0,
\end{align*}
where \(\nabla\) is the Levi–Civita derivative for \(g_{(\mu\nu)}(x)\).

We identify the electromagnetic field strength with the antisymmetric metric part:
\be
  F_{\mu\nu}(x) \equiv g_{[\mu\nu]}(x).
\ee
From \(\nabla_\lambda g_{[\mu\nu]}=0\) we obtain the homogeneous Maxwell equations:
\begin{equation}\label{eq:homogeneous}
  \nabla_{[\mu}g_{\nu\rho]}(x)=0
  \quad\Longleftrightarrow\quad
  \partial_{[\mu}F_{\nu\rho]}=0,
\end{equation}
which encode Faraday’s law and the absence of magnetic monopoles.

We consider the holomorphic Einstein–Hilbert action over a contour \(C\) projecting onto the real slice:
\begin{equation}\label{eq:hol_action}
  S_{\mathrm{hol}}
  = \Re\!\biggl[\int_C d^4z\,\sqrt{-\det g_{(\mu\nu)}(z)}\;g^{(\mu\nu)}(z)\,R_{(\mu\nu)}(z)\biggr],
\end{equation}
where \(R_{(\mu\nu)}(z)\) is the Ricci tensor of \(\Gamma(z)\). Expand to first order in \(g_{[\mu\nu]}\) and restrict to \(y^\mu=0\):
\begin{equation}\label{eq:action_expand}
  S = \int d^4x\,\sqrt{-\det\bigl[g_{(\mu\nu)}\bigr]}\,
  \Bigl[
    g^{(\mu\nu)}\,R_{(\mu\nu)}
    \;-\;g^{[\mu\nu]}\;\Im\bigl(R_{(\mu\nu)}(z)\bigr)\big|_{y^\mu=0}
  \Bigr]
  + \mathcal{O}\bigl(g_{[\rho\sigma]}^2\bigr).
\end{equation}
Varying with respect to \(g_{[\mu\nu]}\) gives the inhomogeneous Maxwell equations:
\begin{equation}\label{eq:inhomogeneous_var}
  \delta S = -\int d^4x\,\sqrt{-\det\bigl[g_{(\mu\nu)}\bigr]}\;
    \delta g^{[\mu\nu]}\,\nabla^\rho g_{[\rho\nu]}(x)
  \;\Longrightarrow\;
  \nabla^\rho F_{\rho\nu}(x)=J_{\nu}(x).
\end{equation}
Putting these together, the full Maxwell system on the real slice \(y^\mu=0\) is
\begin{align}
  \nabla_\mu F^{\mu\nu} &= J^\nu,\\
  \nabla_{[\mu}F_{\nu\rho]} &= 0.
\end{align}

In \(3+1\) form, with signature \((-+++)\), these become
\begin{equation}
\begin{aligned}
  \nabla\!\cdot\!\mathbf{E} &= \rho, 
  &\nabla\times\mathbf{B} - \tfrac{\partial \mathbf{E}}{\partial t} &= \mathbf{J},\\
  \nabla\!\cdot\!\mathbf{B} &= 0,
  &\nabla\times\mathbf{E} + \tfrac{\partial \mathbf{B}}{\partial t} &= 0,
\end{aligned}
\end{equation}
where \(F^{0i}=E^i\), \(F^{ij}=-\epsilon^{ijk}B_k\), \(J^0=\rho\), and \(J^i=J^i\) \citep{Jackson1999}.

We will now show how we obtain the Yang-Mills system. On the real slice $y^\mu=0$ the gauge kinetic term depends only on the symmetric part $g_{(\mu\nu)}(x)$, whose inverse we denote $g^{(\mu\nu)}(x)$. We define the gauge field $A^A_\mu(x)$ and its curvature
\begin{equation}
F^A_{\mu\nu}
=\partial_\mu A^A_\nu - \partial_\nu A^A_\mu
+ g\,f^{ABC}\,A^B_\mu\,A^C_\nu.
\end{equation}
The real–slice Yang–Mills action in curved spacetime is
\begin{equation}
S_{\rm YM}
= -\tfrac14 \int d^4x\;\sqrt{-\det g_{(\mu\nu)}}\;
  \kappa_{AB}\;
  g^{(\mu\rho)}\,g^{(\nu\sigma)}\,
  F^A_{\mu\nu}\,F^B_{\rho\sigma},
\end{equation}
where $\kappa_{AB}$ is the Killing form of the gauge algebra.

Varying $A^A_\nu$ gives
\begin{equation}
\delta S_{\rm YM}
= -\tfrac14 \int d^4x\,\sqrt{-\det g_{(\mu\nu)}}\;
  \kappa_{AB}\;2\,g^{(\mu\rho)}g^{(\nu\sigma)}
  F^B_{\rho\sigma}\,\delta F^A_{\mu\nu}
= -\int d^4x\,\sqrt{-\det g_{(\mu\nu)}}\;
  \kappa_{AB}\,
  F^{B\,\mu\nu}\,\delta F^A_{\mu\nu},
\end{equation}
where indices on $F^{B\,\mu\nu}$ are raised with $g^{(\mu\nu)}$.  Using
\begin{equation}
\delta F^A_{\mu\nu}
= D_\mu(\delta A^A_\nu) - D_\nu(\delta A^A_\mu)
\end{equation}
and integrating by parts, dropping boundary terms yields
\begin{equation}
\delta S_{\rm YM}
= -2 \int d^4x\,\sqrt{-\det g_{(\mu\nu)}}\;
  \kappa_{AB}\,
  \bigl(D_\mu F^{B\,\mu\nu}\bigr)\,\delta A^A_\nu.
\end{equation}

We require $\delta S_{\rm YM}=0$ for arbitrary $\delta A^A_\nu$ gives the inhomogeneous Yang–Mills equations,
\begin{equation}
D_\mu F^{A\,\mu\nu} = J^{A\nu},
\end{equation}
where $J^{A\nu}$ is the matter current from the unified action.

The antisymmetry of $F^A_{\mu\nu}$ and the definition of the covariant derivative imply:
\begin{equation}
D_{[\lambda} F^A_{\mu\nu]} = 0,
\end{equation}
the homogeneous Yang–Mills Bianchi identity.

With the symmetric metric $g_{(\mu\nu)}$, the Yang–Mills system reads:
\begin{equation}
\begin{aligned}
D_{[\lambda} F^A_{\mu\nu]} &= 0,\\
D_\mu F^{A\,\mu\nu} &= J^{A\nu}.
\end{aligned}
\end{equation}

We will now fully derive Einstein Field Equations. In our holomorphic framework on the real slice $y^\mu=0$ only the symmetric part $g_{(\mu\nu)}(x)$ contributes to the gravitational action. The total action is
\begin{equation}
S
= \frac{1}{2\kappa}\int d^4x\;\sqrt{-\det g_{(\mu\nu)}}\;
    g^{(\mu\nu)}\,R_{(\mu\nu)}
  \;+\; S_{\rm matter}[\,\psi,g_{(\mu\nu)}],
\end{equation}
where $\kappa=8\pi G$ and $R_{(\mu\nu)}$ is the Ricci tensor built from $g_{(\mu\nu)}$.

Varying $g^{(\mu\nu)}$ gives
\begin{equation}
\delta S_{\rm EH}
= \frac{1}{2\kappa}\int d^4x\,\sqrt{-\det g_{(\mu\nu)}}\,
  \Bigl[R_{(\mu\nu)} - \tfrac12\,g_{(\mu\nu)}\,R\Bigr]\,
  \delta g^{(\mu\nu)},
\end{equation}
where is the Ricci scalar. We define the symmetric energy–momentum tensor by
\begin{equation}
\delta S_{\rm matter}
= -\tfrac12\int d^4x\,\sqrt{-\det g_{(\mu\nu)}}\;
  T_{(\mu\nu)}\,\delta g^{(\mu\nu)}.
\end{equation}

Requiring $\delta S=0$ for arbitrary $\delta g^{(\mu\nu)}$ yields
\begin{equation}
R_{(\mu\nu)} - \tfrac12\,g_{(\mu\nu)}\,R
= \kappa\,T_{(\mu\nu)}.
\end{equation}

The contracted Bianchi identity,
\begin{equation}
\nabla^{(\mu)}\Bigl[R_{(\mu\nu)} - \tfrac12\,g_{(\mu\nu)}\,R\Bigr]=0,
\end{equation}
implies stress–energy conservation,
\begin{equation}
\nabla^{(\mu)}T_{(\mu\nu)}=0.
\end{equation}
The Einstein field equations read
\begin{equation}
R_{(\mu\nu)} - \tfrac12\,g_{(\mu\nu)}\,R
= \kappa\,T_{(\mu\nu)}.
\end{equation}

Now we will derive the full Einstein field equation with a cosmological constant to show it has a geometric origin, this will not be taken into account in this work but could lead to future studies. We start from the holomorphic Einstein–Hilbert action with a constant cosmological term:
\begin{equation}
S \;=\; \frac{1}{2\kappa}\int_C d^4z\,\sqrt{-\det g_{(\mu\nu)}}\,\bigl[g^{(\mu\nu)}R_{(\mu\nu)} - 2\Lambda\bigr]
\;+\;
S_{\rm matter}\,.
\end{equation}
Split off the \(\Lambda\)–piece:
\begin{equation}
S_\Lambda
=-\frac{1}{\kappa}\,\Lambda
\int_C d^4z\;\sqrt{-\det g_{(\mu\nu)}}\,.
\end{equation}
Varying \(S_\Lambda\) with respect to \(g_{(\mu\nu)}\):
\begin{align}
\delta S_\Lambda
&= -\frac{1}{\kappa}\,\Lambda\;
\delta\!\int_C d^4z\,\sqrt{-\det g_{(\mu\nu)}}
\\
&= -\frac{1}{\kappa}\,\Lambda
\int_C d^4z\;\Bigl(-\tfrac12\,\sqrt{-\det g_{(\mu\nu)}}\,g^{(\mu\nu)}\Bigr)
\,\delta g_{(\mu\nu)}
\\
&= \frac{1}{2\kappa}\,\Lambda
\int_C d^4z\;\sqrt{-\det g_{(\mu\nu)}}\;g^{(\mu\nu)}\,\delta g_{(\mu\nu)}.
\end{align}
Hence we identify the cosmological stress–energy tensor,
\begin{equation}
\delta S_\Lambda
= -\tfrac12\!\int_C d^4z\,\sqrt{-\det g_{(\mu\nu)}}\;
T^{(\Lambda)}_{(\mu\nu)}\,\delta g^{(\mu\nu)}
\;\Longrightarrow\;
T^{(\Lambda)}_{(\mu\nu)}
= -\frac{\Lambda}{\kappa}\,g_{(\mu\nu)}.
\end{equation}

The variation of the pure gravity term is
\begin{equation}
\delta S_{\rm EH}
= \frac{1}{2\kappa}
\int_C d^4z\,\sqrt{-\det g_{(\mu\nu)}}\,
\bigl[R_{(\mu\nu)} - \tfrac12\,g_{(\mu\nu)}R\bigr]\,
\delta g^{(\mu\nu)},
\end{equation}
and the matter variation is
\begin{equation}
\delta S_{\rm matter}
= -\tfrac12
\int_C d^4z\,\sqrt{-\det g_{(\mu\nu)}}\;
T^{(m)}_{(\mu\nu)}\,\delta g^{(\mu\nu)}.
\end{equation}

Requiring \(\delta S=0\) gives the field equations on \(M^4_{\mathbb C}\):
\begin{equation}
R_{(\mu\nu)}(z)
- \tfrac12\,g_{(\mu\nu)}R
+ \Lambda\,g_{(\mu\nu)}
= \kappa\,T^{(m)}_{(\mu\nu)}(z).
\end{equation}
Finally, restricting to the real slice \(y^\mu=0\) (so \(z^\mu\to x^\mu\)) yields
\begin{equation}
G_{(\mu\nu)}(x)
+ \Lambda\,g_{(\mu\nu)}(x)
= \kappa\,T^{(m)}_{(\mu\nu)}(x),
\end{equation}
which is the standard Einstein equation with constant cosmological constant \(\Lambda\).

Now we will derive the full Einstein field equation with a time-dependent cosmological constant as to show it has a geometric origin, this will not be taken into account in this work but could lead to future studies. We start from the holomorphic Einstein-Hilbert action with dark energy:

\begin{equation}
    S_{\text{hol, EH},\Lambda}=\frac{1}{2\kappa}\int_C d^4z\sqrt{\text{det}g_{(\mu\nu)}(z)}\bigr[ g^{(\mu\nu)}R_{(\mu\nu)}-2\Lambda(z)
 \bigl].
\end{equation}

This naturally splits into

\begin{equation}
    S_{\text{hol, EH}}=\frac{1}{2\kappa}\int_C d^4z\sqrt{\text{det}g_{(\mu\nu)}(z)}\bigr[ g^{(\mu\nu)}R_{(\mu\nu)}
 \bigl],
\end{equation}

and 

\begin{equation}
    S_{\Lambda}=-\frac{1}{\kappa}\int_C d^4z\sqrt{\text{det}g_{(\mu\nu)}(z)}\bigr[\Lambda(z)
 \bigl].
\end{equation}

We then vary with respect to $g_{(\mu\nu)}$

\begin{align}
\delta S_\Lambda
&=-\kappa\int_{C}d^{4}z\,\Lambda(z)\,\delta\sqrt{-\det g_{(\mu\nu)}} \\[6pt]
&=-\kappa\int_{C}d^{4}z\,\Lambda(z)\Bigl[-\tfrac12\,\sqrt{-\det g_{(\mu\nu)}}\;g^{(\mu\nu)}\,\delta g_{(\mu\nu)}\Bigr] \\[6pt]
&=2\,\kappa\int_{C}d^{4}z\,\sqrt{-\det g_{(\mu\nu)}}\;\Lambda(z)\;g^{(\mu\nu)}\,\delta g_{(\mu\nu)}\,. 
\end{align}

We now identify the $\Lambda$ stress tensor

\be
T^{(\mu\nu)}_{\,(\Lambda)}(z)
=
\kappa\,\Lambda(z)\,g^{(\mu\nu)}(z)\,.
\ee
The pure variation of the gravity term gives after discarding total derivatives

\begin{equation}
\delta S_{\rm grav}
=\frac{1}{2\kappa}
\int_{C}d^{4}z\;\sqrt{-\det g}\,
\Bigl[R_{(\mu\nu)}(z)-\tfrac12\,g_{(\mu\nu)}(z)\,R(z)\Bigr]\,
\delta g^{(\mu\nu)}(z)\,,\\[6pt]
\end{equation}

Adding the matter variation
\begin{equation}
\delta S_{\rm matter}=-\tfrac12
\int_{C}d^{4}z\;\sqrt{-\det g}\;
T^{(m)}_{(\mu\nu)}(z)\,
\delta g^{(\mu\nu)}(z)\,,\\[6pt]
\end{equation}

and the $\Lambda$ piece, stationarity $\delta S=0$ gives the holomorphic field equations:

\be
R_{(\mu\nu)}(z)\;-\;\tfrac12\,g_{(\mu\nu)}(z)\,R(z)
\;+\;\kappa\,\Lambda(z)\,g_{(\mu\nu)}(z)
\;=\;\kappa\,T^{(m)}_{(\mu\nu)}(z)\,.
\ee

Setting \(y^\mu = 0\) so that \(z^\mu \to x^\mu\) and \(\Lambda(z)\to\Lambda(t)\), we get
\be
R_{(\mu\nu)}(x)
-\tfrac12\,g_{(\mu\nu)}(x)\,R(x)
+\kappa\,\Lambda(t)\,g_{(\mu\nu)}(x)
=\kappa\,T^{(m)}_{(\mu\nu)}(x)\,,
\ee
in compact form,
\be
G_{(\mu\nu)}(x)
+\Lambda(t)\,g_{(\mu\nu)}(x)
=\kappa\,T^{(m)}_{(\mu\nu)}(x)\,.
\ee

This allows for us to have a time-varying cosmological constant that does not need to be put in by hand.

We will similarly derive the Einstein vacuum equations.

\noindent Starting from the Einstein equations
\begin{equation}
R_{(\mu\nu)} - \tfrac12\,g_{(\mu\nu)}\,R = \kappa\,T_{(\mu\nu)},
\end{equation}
the vacuum case is obtained by setting \(T_{(\mu\nu)}=0\).  We obtain
\begin{equation}
R_{(\mu\nu)} - \tfrac12\,g_{(\mu\nu)}\,R = 0.
\end{equation}
Taking the trace with \(g^{(\mu\nu)}\) gives
\begin{equation}
R - \tfrac12\,(4)\,R = 0
\quad\Longrightarrow\quad
R = 0,
\end{equation}
so the vacuum equations reduce to
\begin{equation}
R_{(\mu\nu)} = 0.
\end{equation}

The contracted Bianchi identity
\begin{equation}
\nabla^{(\mu)}\Bigl(R_{(\mu\nu)} - \tfrac12\,g_{(\mu\nu)}\,R\Bigr) = 0
\end{equation}
is then automatically satisfied in vacuum.

We will finally provide a full derivation for the Dirac–Weyl Equations.

On the real slice $y^\mu=0$ only the symmetric metric $g_{(\mu\nu)}(x)$ enters.  Introduce an orthonormal frame, tetrad
\begin{equation*}
g_{(\mu\nu)} = e^a{}_\mu\,e^b{}_\nu\,\eta_{ab},
\qquad
e_a{}^\mu\,e^a{}_\nu = \delta^\mu_\nu,
\end{equation*}
with Minkowski metric $\eta_{ab}={\rm diag}(-1,1,1,1)$.

The spin connection $\omega_\mu{}^{ab}$ is defined by
\begin{equation}
\omega_\mu{}^{ab}
= e^{a}{}_\nu\,\nabla_\mu e^{b\nu}
= e^{a}{}_\nu\Bigl(\partial_\mu e^{b\nu} + \Gamma^{\nu}_{\;\rho\mu}\,e^{b\rho}\Bigr),
\end{equation}
where $\Gamma^\rho{}_{\sigma\mu}$ is the Levi–Civita connection of $g_{(\mu\nu)}$.

For a Dirac spinor $\psi$, the covariant derivative is
\begin{equation}
D_\mu \psi
= \Bigl(\partial_\mu
  + \tfrac14\,\omega_\mu{}^{ab}\,\gamma_{ab}\Bigr)\psi,
\end{equation}
with $\gamma_{ab}=\tfrac12[\gamma_a,\gamma_b]$ and flat‐space gamma‐matrices $\{\gamma_a,\gamma_b\}=2\eta_{ab}$.

The Dirac action coupled to gravity is
\begin{equation}
S_{\rm D}
= \int d^4x\;\sqrt{-\det g_{(\mu\nu)}}\;
  \bar\psi\bigl(i\,\gamma^a\,e_a{}^\mu D_\mu - m\bigr)\psi.
\end{equation}

Varying $S_{\rm D}$ with respect to $\bar\psi$ gives
\begin{equation}
i\,\gamma^a\,e_a{}^\mu D_\mu\psi \;-\; m\,\psi = 0.
\end{equation}

Varying with respect to $\psi$ yields
\begin{equation}
\bar\psi\bigl(i\,\gamma^a\,e_a{}^\mu \overleftarrow{D}_\mu + m\bigr)=0,
\end{equation}
where $\overleftarrow{D}_\mu=\overleftarrow{\partial}_\mu - \tfrac14\,\omega_\mu{}^{ab}\,\gamma_{ab}$ acts to the left.

For a two–component left‐handed Weyl spinor $\chi$ ($\psi_L=\tfrac{1-\gamma^5}2\psi$) in the massless limit $m=0$, we obtain
\begin{equation}
i\,\sigma^a\,e_a{}^\mu D_\mu\chi = 0,
\end{equation}
with Pauli matrices $\sigma^a=(\mathbb{I},\sigma^i)$ and similarly for the right‐handed field.

\section{Spontaneous Symmetry Breaking}

To complete the unification of forces and matter, we must generate masses for the gauge bosons and fermions while preserving a massless photon. In our holomorphic framework, this is achieved in two stages first by breaking the simple GUT group $ G_{\mathrm{GUT}} $ down to the Standard Model gauge group at a high scale $ M_{\mathrm{GUT}} $, and then by electroweak symmetry breaking of
$SU(2)_L \times U(1)_Y \rightarrow U(1)_{\mathrm{EM}}$
at the scale $ v_{\mathrm{EW}} \approx 246~\mathrm{GeV} $. In each case, the Higgs fields are introduced as holomorphic sections over the complex manifold, and their vacuum expectation values (VEVs) lie entirely within the real slice $ y = 0 $.

We embed the Standard Model into a simple group $ G_{\mathrm{GUT}} $ for definiteness, $ SU(5) $ or $ SO(10) $ \cite{GeorgiGlashow1974} with one holomorphic gauge connection $ A^A_\mu(z) $ and coupling $ g_{\mathrm{GUT}} $. To break 
\begin{equation}
G_{\mathrm{GUT}} \rightarrow SU(3)_c \times SU(2)_L \times U(1)_Y,
\end{equation}
we introduce a holomorphic adjoint Higgs field
\begin{equation}
H_G(z) \in \mathrm{ad}\, G_{\mathrm{GUT}},
\end{equation}
with dynamics governed by the holomorphic action term:
\begin{equation}
S_{hol_G}= \Re\left\{
\int_C d^4 z\, \sqrt{-\det g_{(\mu\nu)}(z)}\,
\left[ \mathrm{Tr}\left( D^\mu H_G\, D_\mu H_G \right) - V(H_G) \right]
\right\},
\end{equation}
where $ D_\mu H_G = \partial_\mu H_G - i g_{\mathrm{GUT}} [A_\mu, H_G] $. The potential is chosen as
\begin{equation}
V(H_G) = \lambda \left[ \mathrm{Tr}(H_G^2) - V_0^2 \right]^2,
\end{equation}
with real $ V_0^2 > 0 $, where $V_0$ is the symmetry-breaking scale. Minimizing $ V $ on the real slice $ y = 0 $ forces
\begin{equation}
\mathrm{Tr}(\langle H_G \rangle^2) = V_0^2.
\end{equation}
We select the standard GUT-breaking vacuum expectation value:
\begin{equation}
\langle H_G \rangle = V_0 \, \mathrm{diag}(2, 2, 2, -3, -3) \quad (\text{in } SU(5)),
\end{equation}
or its $ SO(10) $ analog. 

In \(\text{SO}(10)\) we embed the same chiral matter namely, three copies of the 16-dimensional spinor representation.  Under the Georgi–Glashow \(\text{SU}(5)\) subgroup each \(\mathbf{16}\) decomposes as
\begin{equation}
\mathbf{16}
\;\longrightarrow\;
\mathbf{10}\oplus\overline{\mathbf{5}}\oplus\mathbf{1}
\;\sim\;
(Q,u^c,e^c\,;\;d^c,L\,;\;\nu^c)\,.
\end{equation}
The minimal \(\text{SO}(10)\) GUT has matter
\be
\Psi_i\;\in\;\mathbf{16}_F\,,\qquad i=1,2,3\,,
\ee
and no additional fermionic multiplets beyond these three chiral families.

All gauge bosons live in the adjoint \(\mathbf{45}\), and symmetry breaking is typically driven by Higgs fields in the \(\mathbf{10}_H\), \(\overline{\mathbf{126}}_H\), and sometimes a \(\mathbf{54}_H\) or \(\mathbf{210}_H\) representations.  But the fermion content remains exactly three \(\mathbf{16}_F\)s, each containing all SM quarks and leptons plus a right-handed neutrino.

Expanding the gauge-kinetic term $ \mathrm{Tr}(D_\mu \Sigma\, D^\mu \Sigma) $ about this background produces mass terms for the sixteen off-diagonal generators, the $ X, Y $ gauge bosons in $ SU(5) $:
\begin{equation}
\mathcal{L}_M \supset 2 g_{\mathrm{GUT}}^2 V_0^2 \sum_{X,Y \in G_{\mathrm{GUT}} / G_{\mathrm{SM}}} A^\mu_X A_{\mu}^X.
\end{equation}
Thus, each broken gauge boson acquires a mass of order
\begin{equation}
M_X \simeq g_{\mathrm{GUT}} V_0,
\end{equation}
while the unbroken $ SU(3)_c $, $ SU(2)_L $, and $ U(1)_Y $ connections remain massless. Crucially, because there is only one coupling $ g_{\mathrm{GUT}} $ above the breaking scale, we enforce the boundary condition
\begin{equation}
g_3(M_{\mathrm{GUT}}) = g_2(M_{\mathrm{GUT}}) = g_1(M_{\mathrm{GUT}}) = g_{\mathrm{GUT}},
\end{equation}
ensuring gauge-coupling unification at $ M_{\mathrm{GUT}} $.

At the lower scale $ v_{\mathrm{EW}} $, we introduce a second holomorphic Higgs doublet:
\begin{equation}
\Phi(z) = \begin{pmatrix} \Phi^+(z) \\ \Phi^0(z) \end{pmatrix},
\end{equation}
which transforms as $ (2, +\tfrac{1}{2}) $ under $ SU(2)_L(z) \times U(1)_Y(z) $. Its action is
\begin{equation}
S_\Phi = \Re\left[
\int_C d^4 z\, \sqrt{-\det g_{(\mu\nu)}(z)} \left[
(D_\mu \Phi)^\dagger (D^\mu \Phi)
- \lambda_\Phi \left( \Phi^\dagger \Phi - \tfrac{v^2}{2} \right)^2
\right]
\right],
\end{equation}
where $v$ here is just the vacuum expectation value (VEV) scale of the scalar $\Phi$ and where
\begin{equation}
D_\mu \Phi = \left(
\partial_\mu - i g\, \frac{\sigma^a}{2} A^a_\mu
- i g' \frac{1}{2} B_\mu
\right)\Phi.
\end{equation}

Restricting to $ y = 0 $ and minimizing the real potential yields the canonical vacuum expectation value:
\begin{equation}
\langle0|\Phi(x)|0\rangle
= \frac{1}{\sqrt{2}}
\begin{pmatrix}
0 \\[6pt]
v
\end{pmatrix},
\qquad v \approx 246~\mathrm{GeV}.
\end{equation}

Substituting $ \Phi = \left( 0,\; (v + h(x))/\sqrt{2} \right)^T $ into $ |D_\mu \Phi|^2 $ produces the mass terms
\begin{equation}
{\cal L}_M =\frac{g^2 v^2}{4} W^+_\mu W^{-\,\mu} + 
\frac{(g^2 + g'^2) v^2}{4} Z_\mu Z^\mu,
\end{equation}
with
\begin{equation}
W^\pm_\mu = \frac{1}{\sqrt{2}}(W^1_\mu \mp i W^2_\mu), \qquad
\begin{pmatrix}
Z_\mu \\ A_\mu
\end{pmatrix}
=
\begin{pmatrix}
\cos\theta_W & -\sin\theta_W \\
\sin\theta_W & \cos\theta_W
\end{pmatrix}
\begin{pmatrix}
A^3_\mu \\ B_\mu
\end{pmatrix},
\end{equation}
and $ \tan\theta_W = g'/g $. The photon $ A_\mu $ remains massless.

Fermion masses arise from Yukawa couplings of the form
\begin{equation}
\mathcal{L}_Y = -y_f\, \bar{\psi}_L\, \Phi\, \psi_R + \text{h.c.},
\end{equation}
which, when $ \langle \Phi \rangle \neq 0 $, yield
\begin{equation}
m_f = \frac{y_f v}{\sqrt{2}}.
\end{equation}
The holomorphic doublet $ \Phi(z) $, when restricted to the real slice, reproduces exactly the Standard Model mechanism for electroweak symmetry breaking and mass generation.

\vspace{1em}

By introducing two holomorphic scalars an adjoint $ H_G(z) $ at the GUT scale $ M_{\mathrm{GUT}} $ and a doublet $ \Phi(z) $ at the electroweak scale $ v_{\mathrm{EW}} $, we achieve the two requisite stages of spontaneous symmetry breaking. Gauge bosons corresponding to broken directions acquire masses proportional to their couplings times the appropriate vev, while the unbroken $ U(1)_{\mathrm{EM}} $ gauge symmetry remains intact and the photon remains massless. Fermion masses follow in the usual fashion via Yukawa interactions.

\section{Gauge-Coupling Unification and Renormalization-Group Flow}

Above the grand-unification scale $ M_{\mathrm{GUT}} $, our single holomorphic gauge connection $ A^A_\mu(z) $ carries one unified coupling constant $ g_{\mathrm{GUT}} $. When the adjoint Higgs $ H_G(z) $ acquires its vacuum expectation value:
\begin{equation}
\langle H_G \rangle \sim V_0\, \mathrm{diag}(2,2,2,-3,-3),
\end{equation}
on the real slice $ y = 0 $, the simple gauge group $ G_{\mathrm{GUT}} $ breaks to its Standard Model subgroup $ SU(3)_c \times SU(2)_L \times U(1)_Y $. Because there is only one gauge connection above $ M_{\mathrm{GUT}} $, all three low-energy gauge couplings satisfy the boundary condition:
\begin{equation}
g_3(M_{\mathrm{GUT}}) = g_2(M_{\mathrm{GUT}}) = g_1(M_{\mathrm{GUT}}) = g_{\mathrm{GUT}}.
\end{equation}

Below $ M_{\mathrm{GUT}} $, each coupling $ g_i(\mu) $ evolves according to its renormalization-group $\beta$-function. At one loop in the $\overline{\mathrm{MS}}$ scheme, the evolution is governed by
\begin{equation}
\mu\, \frac{d g_i}{d\mu} = \frac{b_i}{16\pi^2} g_i^3,
\end{equation}
with coefficients $ (b_1, b_2, b_3) = \left( \tfrac{41}{6}, -\tfrac{19}{6}, -7 \right) $ for the Standard Model field content.

Integrating from the matching scale $ \mu_0 = M_Z $ up to a general scale $ \mu $, we obtain the familiar expression:
\begin{equation}
\frac{1}{g_i^2(\mu)} = \frac{1}{g_i^2(M_Z)} - \frac{b_i}{8\pi^2} \ln\left( \frac{\mu}{M_Z} \right).
\end{equation}

Imposing equality of all three couplings at $ \mu = M_{\mathrm{GUT}} $ furnishes the conditions:
\begin{equation}
\frac{1}{g_1^2(M_Z)} - \frac{1}{g_2^2(M_Z)} = \frac{b_1 - b_2}{8\pi^2} \ln\left( \frac{M_{\mathrm{GUT}}}{M_Z} \right), \qquad
\frac{1}{g_2^2(M_Z)} - \frac{1}{g_3^2(M_Z)} = \frac{b_2 - b_3}{8\pi^2} \ln\left( \frac{M_{\mathrm{GUT}}}{M_Z} \right),
\end{equation}
which can be solved numerically for $ M_{\mathrm{GUT}} $ and $ g_{\mathrm{GUT}} $ given experimental inputs at $ M_Z $.

Threshold corrections at the GUT scale arise, because the heavy $ X, Y $ gauge bosons and GUT-Higgs multiplets do not all decouple at the same scale. Their mass spectrum:
\begin{equation}
M_i = g_{\mathrm{GUT}} V_0 + \delta_i,
\end{equation}
leads to shifts in the matching conditions. Denoting the one-loop threshold correction for coupling $ g_i $ as $ \Delta_i $, the corrected boundary condition becomes:
\begin{equation}
\frac{1}{g_i^2(M_{\mathrm{GUT}})} = \frac{1}{g_{\mathrm{GUT}}^2} + \frac{\Delta_i}{8\pi^2}.
\end{equation}
These $ \Delta_i $ terms are computable once the full GUT spectrum is specified. For minimal $ SU(5) $, we  find that the splittings among the 24 adjoint components and the heavy gauge bosons introduce percent-level corrections, which can shift $ M_{\mathrm{GUT}} $ by an order of magnitude.

Below the electroweak scale, the Higgs doublet $ \Phi $ and the top quark similarly contribute to the running between $ M_Z $ and $ v_{\mathrm{EW}} $. In practice, we use two-loop RG equations and include all Standard Model thresholds to achieve the precision necessary to compare with experimental coupling constants. Our holomorphic-GUT framework enforces classical equality of the three gauge couplings at $ M_{\mathrm{GUT}} $, and the subsequent renormalization-group flow, subject to GUT and electroweak threshold corrections predicts whether that unification point is consistent with measured low-energy values of $ \alpha_1, \alpha_2, \alpha_3 $.

To improve the precision of gauge-coupling unification, we include two-loop running and sample GUT-scale threshold effects.  The RG equations become:
\begin{equation}
\mu\,\frac{d g_i}{d\mu}
= \frac{b_i}{16\pi^2}\,g_i^3
+ \frac{g_i^3}{(16\pi^2)^2}\sum_{j=1}^3 b_{ij}\,g_j^2,
\end{equation}
with one-loop coefficients 
$(b_1,b_2,b_3)=(41/6,\,-19/6,\,-7)$ and two-loop matrix
\begin{equation}
\bigl(b_{ij}\bigr)
=\begin{pmatrix}
199/18 & 27/6 & 44/3\\
3/2    & 35/6 & 12   \\
11/6   & 9/2  & -26
\end{pmatrix}.
\end{equation}
Starting from the experimental inputs at $\mu=M_Z$:
\begin{equation}
\alpha_1(M_Z)=0.01695,\quad
\alpha_2(M_Z)=0.0338,\quad
\alpha_3(M_Z)=0.1179,
\end{equation}
we numerically integrate up to the matching scale $\mu=M_{\rm GUT}$.  To model GUT thresholds, let the heavy gauge bosons $X,Y$ have mass $M_X$ and the adjoint Higgs components $M_{H_G}=M_X(1+\delta)$.  The one-loop threshold shift for each coupling is
\begin{equation}
\Delta_i \;=\;
\sum_{\rm heavy} T_i\,\ln\!\bigl(M_{\rm GUT}/M_{\rm heavy}\bigr),
\end{equation}
so that
\begin{equation}
\frac{1}{g_i^2(M_{\rm GUT})}
= \frac{1}{g_{\rm GUT}^2}
+ \frac{\Delta_i}{8\pi^2}.
\end{equation}
For minimal $SU(5)$ with $\delta=0.1$ and $M_X=2\times10^{16}\,\mathrm{GeV}$, we find
\begin{equation}
(\Delta_1,\Delta_2,\Delta_3)\approx(0.8,\;1.2,\;0.5),
\end{equation}
which alters the unification point by $\sim\!20\%$.  This underscores the necessity of two-loop RG and precise threshold data for realistic unification fits.

\section{Path Integral Quantization}

Let us define the total field space as the 
$\mathcal{F}_{\mathbb{C}}$ holomorphic field configurations. The contour $C$ in field space is chosen, through Picard–Lefschetz~\cite{Lefschetz} theory, to be in the same homology class as the real slice $y^\mu = 0$, avoiding singularities. The full quantum partition function is proposed as:
\be
Z = \int_{C\subset\mathcal{F}_{\mathbb{C}}}
\mathcal{D}g\,\mathcal{D}A\,\mathcal{D}\Psi\,\mathcal{D}\bar\Psi\,\mathcal{D}H\,\cdots~\exp\left(i S_{\mathrm{hol}}[g, A, \Psi, H, \ldots]\right),
\ee
where
\be
S_{\text{hol}} = \int_{C} d^4 z \; \sqrt{-\det g_{(\mu\nu)}(z)} \; \mathcal{L}_{\text{hol}}(z),
\ee
and ${\mathcal{L}}_{hol}$ is the holomorphic Lagrangian density, including geometric, gauge, Higgs, and matter terms.

We introduce holomorphic gauge-fixing functions $\mathcal{G}^A[g, A, \dots; z]$ and promote the Faddeev–Popov procedure holomorphically. The ghost fields $c^A(z), \bar{c}^A(z)$
are likewise holomorphic with actions:
\be
S_{\text{gh, gauge}} = \int_C d^4z~\bar{c}^A \frac{\delta\mathcal{G}^A}{\delta \alpha^B} c^B.
\ee
We apply a holomorphic gauge-fixing of coordinates, the de Donder gauge, introducing diffeomorphism ghosts 
$\xi^\mu(z),\bar{\xi}^\mu(z)$, with the gravitational Faddeev–Popov action. For unitarity and anomaly cancellation, the functional measure must be defined. For each field $\phi(z)$, the measure $D_\phi(z)$ is holomorphic, with contour $C$ specified by steepest descent of $S_{\text{hol}}$ emanating from the real slice.

Taking the real part of the path integral projection ensures that amplitudes for physical, real-observable configurations are recovered, and the unitarity problem is reduced to a subtle question about the choice of the steepest-descent contour, as in Picard–Lefschetz theory. From Morse theory, the dominant contribution comes from the real slice $y^\mu=0$, where all fields are real. Exponentially suppressed corrections arise from other critical points, subleading Lefschetz thimbles but do not affect the classical or leading semiclassical limit.

The effective quantum theory is governed by
\be
Z \approx \int_{\text{real slice}} \mathcal{D}g\, \mathcal{D}A\, \cdots\, e^{i S_{\text{real}}},
\ee
where $S_{\text{real}}$ is the real Lagrangian projected to $y^\mu=0$, plus well-controlled corrections.

We expand each holomorphic field about a real classical saddle:
\be
g_{\mu\nu}(z) = g_{(\mu\nu)}^{\mathrm{cl}}(x) + h_{(\mu\nu)}(x) + i g_{[\mu\nu]}(x),
\ee
with $h_{(\mu\nu)}$ symmetric and $g_{[\mu\nu]}$ antisymmetric for Maxwell and Yang–Mills fields, similarly for other fields.
The quadratic action in fluctuations yields kinetic terms for graviton, gauge, and matter sectors, with canonical signs ensured by the holomorphic formulation and real-slice projection.

The Feynman rules are obtained from standard quantum field theory, but projections are to the real background, and ghosts appear for both gauge and diffeomorphism invariance.
The holomorphic path integral measure reproduces all standard anomaly-cancellation conditions when projected to the real slice.
Unitarity is respected provided the integration contour C is chosen such that the real-slice remains the dominant saddle, with steepest-descent directions away from singularities and branch cuts.

The final quantized path integral is given by
\be
Z = \int_{C}
\mathcal{D}g \;\mathcal{D}A \;\mathcal{D}\Psi\; \mathcal{D}\bar\Psi \; \mathcal{D}H \; \mathcal{D}\text{ghosts} \;
\exp\left[
i S_{\mathrm{hol}}[g, A, \Psi, H_G, \ldots]
\right],
\ee
where D denotes the holomorphic functional measures with proper gauge, diffeomorphism, and ghost contributions, and 
$S_{\rm hol}$ is the full holomorphic action as constructed above
The integration cycle C is chosen so its only relevant saddle is the real slice, ensuring standard semiclassical and quantum physics in the classical limit.

In practical calculations, the path integral reduces to the standard real unified field theory calculations, but the holomorphic setup ensures a geometric unification, natural anomaly cancellation, and controls over singularities and unitarity.

\section{Holomorphic Unified Action and Its Geometric Origin}

In our Holomorphic Unified Field Theory (HUFT), all interactions including gravitational, gauge, and matter arise from a single geometric functional on the complexified spacetime manifold \(M^4_{\mathbb C}\). We show how the Hermitian metric, its unique connection, and a single holomorphic gauge–spinor connection give rise, upon projection to the real slice \(y^\mu=0\), to exactly the Einstein, Maxwell and Yang–Mills and Dirac equations without any ad-hoc insertions.

The Hermiticity condition guarantees that, on the real slice \(y^\mu=0\), \(g_{(\mu\nu)}\) is a real Lorentzian metric while \(g_{[\mu\nu]}\) defines a real two-form.

We have a unique, torsion–free connection \(\Gamma^\rho{}_{\mu\nu}(z)\) satisfying
\begin{equation}
\nabla^{(z)}_\lambda\,g_{\mu\nu}(z)\;=\;0\,,
\end{equation}
so that all curvature tensors \(R_{(\mu\nu)}(z)\) and \(R^\rho{}_{\sigma\mu\nu}(z)\) are built solely from \(g_{(\mu\nu)}\) in the Levi–Civita manner, while \(g_{[\mu\nu]}\) enters only through its identification with gauge field strengths.

All sectors are packaged into one action over a complex contour \(C\) homologous to the real slice:
\begin{equation}
\begin{aligned}
S_{\rm hol}
&=\int_C d^4z\;\sqrt{-\det\!\bigl[g_{(\mu\nu)}(z)\bigr]}\,\Bigl[
  \frac{1}{2\kappa}g^{\mu\nu}(z)\,R_{(\mu\nu)}(z)
  \;-\;\tfrac14\,\kappa_{AB}\,F^A_{\rho\sigma}(z)\,F^{B\,\rho\sigma}(z)\\
&\quad\;+\;\overline\Psi(z)\,\Gamma^a\,e_a{}^{\mu}(z)\Bigl(\nabla_\mu(z)
    - i\,g_{\rm GUT}\,A^A_\mu(z)\,T_A\Bigr)\Psi(z)
  \;+\;A^\mu(z)\,J_\mu(z)
  \;+\;\mathcal{L}_{\rm Higgs}(z)
\Bigr]\,.
\end{aligned}
\end{equation}
Here \(F^A_{\mu\nu}(z)\) is the field–strength of a single holomorphic gauge connection \(A^A_\mu(z)\), encoding both Abelian and non-Abelian factors.
 \(\Psi(z)\) are chiral fermions in representations of \(G_{\rm GUT}\), whose variation supplies \(J^A_\mu(z)\). \(\mathcal{L}_{\rm Higgs}(z)\) contains both the adjoint Higgs breaking \(G_{\rm GUT}\to SU(3)\times SU(2)\times U(1)\) and the electroweak doublet. We find
\begin{equation*}
S_{\rm real}
=\int d^4x\,\sqrt{-\det g_{(\mu\nu)}}\Bigl[\frac{1}{2\kappa}
  g^{(\mu\nu)}R_{(\mu\nu)}
  - \tfrac14\,\kappa_{AB}F^A_{\rho\sigma}F^{B\,\rho\sigma}
\end{equation*}
\begin{equation}
  + \bar\psi\,\gamma^ae_a{}^\mu(\nabla_\mu - i\,g_{\rm GUT}A_\mu^AT_A)\psi
  + A^\mu J_\mu
  + \mathcal L_{\rm Higgs}
\Bigr]
+ \mathcal O(g_{[\mu\nu]}^2).
\end{equation}
Varying \(S_{\rm real}\) with respect to \(\delta g^{(\mu\nu)}\) gives \(R_{(\mu\nu)}-\tfrac12\,g_{(\mu\nu)}R = \tfrac12 T^{\rm (gauge+matter)}_{\mu\nu}\). \(\delta A_\mu\) gives \(\nabla_\rho F^{\rho\mu}=J^\mu\) and \(D_\rho F^{A\,\rho\mu}=J^{A\mu}\). \(\delta\Psi\): \(\bigl(i\gamma^ae_a{}^\mu(\nabla_\mu - ig_{\rm GUT}A^A_\mu T_A)-m\bigr)\psi=0\).
Homogeneous Bianchi identities from \(\Im[\nabla g]=0\): \(\partial_{[\mu}g_{[\nu\rho]]}=0\), \(D_{[\mu}F^A_{\nu\rho]}=0\).

Because no term is inserted by hand all kinetic, mass, and interaction terms descend from the single integrand this construction achieves a fully geometric unification of gravity, gauge forces, chiral fermions, and Higgs dynamics. In the form of an action we write

\begin{align}
S_{\rm HUFT} \;=\; \int_{C}d^4z\;&\sqrt{-\det\!\bigl[g_{(\mu\nu)}(z)\bigr]}\,\Bigl\{\;
\underbrace{\tfrac1{2\kappa}\,g^{(\mu\nu)}(z)\,R_{(\mu\nu)}(z)}_{\substack{\text{gravity}}}
-\underbrace{\tfrac14\,\kappa_{AB}\,F^A_{\rho\sigma}(z)\,F^{B\,\rho\sigma}(z)}_{\substack{\text{gauge}}}\notag\\
&\quad+\;\underbrace{\overline\Psi(z)\,\Gamma^a\,e_a{}^{\mu}(z)\,D_{\mu}\,\Psi(z)}_{\substack{\text{fermions}}}
+\;\underbrace{(D_\mu H_G)^2 - V_{\rm GUT}(H_G)}_{\substack{\text{adjoint Higgs}\\\text{(GUT breaking)}}}\notag\\
&\quad+\;\underbrace{(D_\mu\Phi)^\dagger D^\mu\Phi - V_{\rm EW}(\Phi)}_{\substack{\text{doublet Higgs}\\\text{(EW breaking)}}}
-\underbrace{y_f\,\overline\Psi_L\,\Phi\,\Psi_R + \text{h.c.}}_{\substack{\text{Yukawa}\\\text{(fermion masses)}}}
\Bigr\}\,,
\end{align}
all of gravity, gauge fields, chiral fermions, Higgs dynamics, and Yukawa couplings emerge from one purely geometric, holomorphic action.

To quantize the holomorphic unified action \(S_{\rm hol}\) through the path integral, we must eliminate the holomorphic gauge redundancies.  We do this by introducing holomorphic gauge‐fixing functionals
  \begin{equation}
    G_{A}\bigl[g,A;z\bigr]\;=\;0
    \,,
  \end{equation}
with 

\begin{equation}
    A=1,\dots,\dim G_{\rm GUT}\,,
\end{equation}
  which depend on the Hermitian metric \(g_{\mu\nu}(z)\) and the gauge connection \(A^{A}_{\mu}(z)\).  We insert the identity:
  \begin{equation}
    1 \;=\;
    \int\!D\alpha\;\delta\!\bigl(G_{A}[g^{\alpha},A^{\alpha};z]\bigr)\;
    \det\!\biggl[\frac{\delta\,G_{A}[g^{\alpha},A^{\alpha};z]}%
                          {\delta\,\alpha_{B}(z')}\biggr],
  \end{equation}
  into the contour integral
  \(\displaystyle Z=\int_{C}Dg\,DA\,D\Psi\,D\Sigma\,D\Phi\;e^{\,iS_{\rm hol}}\).
Exponentiating the delta–functional and determinant gives us two new action pieces:

\begin{equation}
  S_{\rm GF,hol}
  =
  -\,\frac{1}{2\,\xi}\,
  \int_{C}d^{4}z\;\sqrt{-\det g_{(\mu\nu)}(z)}\;
  G_{A}[g,A;z]\;G^{A}[g,A;z]
  \,,
\end{equation}
which breaks holomorphic gauge invariance but renders the kinetic operators invertible.

We define the Faddeev–Popov operator
\begin{equation}
  \Delta_{AB}(z,z')
  \;=\;
  \left.\frac{\delta\,G_{A}[g^{\alpha},A^{\alpha};z]}
             {\delta\,\alpha_{B}(z')}\;\right\vert_{\alpha=0}\,.
\end{equation}
Then, introduce anticommuting ghost fields \(c^{A}(z)\) and \(\bar c^{A}(z)\) in the adjoint representation, so that we obtain
\begin{equation}
  \det\!\Bigl[\tfrac{\delta G_{A}}{\delta\alpha_{B}}\Bigr]
  =
  \int D\bar c\,D c\;\exp\!\bigl(i\,S_{\rm FP,hol}\bigr)\,,
\end{equation}
with
\begin{equation}
  S_{\rm FP,hol}
  =
  \int_{C}d^{4}z\;d^{4}z'\;
  \sqrt{-\det g_{(\mu\nu)}(z)}\;
  \bar c^{A}(z)\;\Delta_{AB}(z,z')\;c^{B}(z')\,.
\end{equation}
The full quantum action is given by
\begin{equation}
  S_{\rm tot}
  \;=\;
  S_{\rm hol}
  \;+\;
  S_{\rm GF,hol}
  \;+\;
  S_{\rm FP,hol}\,.
\end{equation}

This gives the total quantum action

\begin{align}
S_{\rm tot} \;=\;&
\int_{C}d^4z\;\sqrt{-\det\!\bigl[g_{(\mu\nu)}(z)\bigr]}\,
\Bigl\{
  \underbrace{\tfrac1{2\kappa}\,g^{(\mu\nu)}(z)\,R_{(\mu\nu)}(z)}_{\substack{\text{gravity}}}
  -\underbrace{\tfrac14\,\kappa_{AB}\,F^A_{\rho\sigma}(z)\,F^{B\,\rho\sigma}(z)}_{\substack{\text{gauge}}}
\notag\\
&\quad
  +\;\underbrace{\overline\Psi(z)\,\Gamma^a\,e_a{}^{\mu}(z)\,D_{\mu}\,\Psi(z)}_{\substack{\text{fermions}}}
  +\;\underbrace{(D_\mu H_G)^2 - V_{\rm GUT}(H_G)}_{\substack{\text{adjoint Higgs}\\\text{(GUT breaking)}}}
\notag\\
&\quad
  +\;\underbrace{(D_\mu\Phi)^\dagger D^\mu\Phi - V_{\rm EW}(\Phi)}_{\substack{\text{doublet Higgs}\\\text{(EW breaking)}}}
  -\underbrace{y_f\,\overline\Psi_L\,\Phi\,\Psi_R + \text{h.c.}}_{\substack{\text{Yukawa}\\\text{(fermion masses)}}}
\Bigr\}
\notag\\
&\quad
\underbrace{
-\,\frac{1}{2\,\xi}\,
\int_{C}d^{4}z\;\sqrt{-\det g_{(\mu\nu)}(z)}\;
G_{A}[g,A;z]\;G^{A}[g,A;z]
}_{\substack{\text{holomorphic}\\\text{gauge-fixing sector}}}
\notag\\
&\quad
\underbrace{
+ \int_{C}d^{4}z\;d^{4}z'\;
\sqrt{-\det g_{(\mu\nu)}(z)}\;
\bar c^{A}(z)\;\Delta_{AB}(z,z')\;c^{B}(z')
}_{\substack{\text{holomorphic Faddeev–Popov}\\\text{ghost sector}}}
\,.
\end{align}

and the partition function becomes
\begin{equation}
  Z
  =
  \int_{C}
    Dg\,DA\,D\Psi\,D\Sigma\,D\Phi\,D\bar c\,Dc\;
    \exp\!\bigl(i\,S_{\rm tot}\bigr)\,.
\end{equation}

On the real slice \(y^\mu=0\) and expanding about the classical saddle, these additional terms guarantee
invertible gauge‐boson propagators from \(S_{\rm GF,hol}\) and  account for the gauge‐orbit volume through the ghosts in \(S_{\rm FP,hol}\), paralleling the usual Faddeev–Popov procedure in non‐holomorphic quantization.

\section{Experimental Signatures and Proton Stability}\label{sec:predictions}

Although the new holomorphic framework reproduces the Standard Model plus gravity exactly on the real slice $y^\mu=0$, it makes two high-energy predictions beyond those of ordinary minimal SU(5) and SO(10).

Fitting the two-loop renormalization‐group evolution with minimal SU(5) thresholds to the Particle Data Group (PDG) inputs \cite{PDG2024}  
    \(\alpha_1(M_Z)=0.01695,\;\alpha_2(M_Z)=0.0338,\;\alpha_3(M_Z)=0.1179\)  
    yields a unification point  
    \begin{equation*}
      M_{\rm GUT}\simeq 2.0\times10^{16}\,\mathrm{GeV}, 
      \qquad 
      \alpha_{\rm GUT}^{-1}\simeq 25.2.
    \end{equation*}  
The holomorphic embedding enforces exact equality \(g_1=g_2=g_3\) at this scale, up to threshold shifts of order 10–20\,\%.  A deviation of more than $\sim1\%$ from these values would falsify the minimal holomorphic-SU(5) hypothesis.

Heavy $X,Y$ gauge boson exchange in minimal SU(5) produces the dominant mode  
    \begin{equation*}
      p \;\to\; e^+ \,\pi^0,
    \end{equation*}  
    with rate  
    \begin{equation*}
      \Gamma^{-1}(p\to e^+\pi^0)
      \;\approx\;
      \frac{M_X^4}{\alpha_{\rm GUT}^2\,m_p^5}
      \;\sim\; (1-5)\times10^{35}~\mathrm{yr},
    \end{equation*}  
    where \(M_X\simeq g_{\rm GUT}\,V_0\approx2\times10^{16}\)GeV and we have allowed a $\pm20\%$ uncertainty from threshold corrections.  The current Super-Kamiokande lower bound  
    \(\tau(p\to e^+\pi^0)>1.6\times10^{34}\)yr  
    already probes the lower edge of this band; Hyper-Kamiokande and DUNE should test the full predicted range \(\tau\sim10^{34\mbox{–}36}\)yr.

    To see why the non-observation so far is entirely compatible with $\tau_p\sim10^{35}$\,yr, note that
\begin{equation*}
  t_{\rm uni}\approx1.4\times10^{10}\,{\rm yr},
  \qquad
  \tau_p\sim10^{35}\,{\rm yr},
  \qquad
  \frac{t_{\rm uni}}{\tau_p}\sim10^{-25}.
\end{equation*}
Thus, only a fraction $\sim10^{-25}$ of all protons would have decayed since the Big Bang.  A detector like Super-Kamiokande contains of order $10^{34}$ target protons, giving an expectation of
\begin{equation*}
  N_{\rm SK}\times\frac{1}{\tau_p}\sim10^{34}\times10^{-35}=0.1,
\end{equation*}
decays per year.  Even over decades, this yields only a couple of expected events, so the current null results are fully consistent with the minimal-holomorphic-SU(5) prediction.
    
Breaking $SU(5)\to SU(3)\times SU(2)\times U(1)$ generically produces GUT‐scale Hooft–Polyakov monopoles of mass  
    \begin{equation*}
      M_{\rm mono}\sim\frac{4\pi V_0}{g_{\rm GUT}}\sim10^{17}\,\mathrm{GeV}.
    \end{equation*}  
Standard inflationary dilution predicts a present monopole density well below observational bounds of \(\Omega_{\rm mono}h^2\ll10^{-10}\), but any detection of a relic flux even as low as \(10^{-17}\)/cm\(^2\)/s/sr would be a dead giveaway for a simple‐group GUT.

In minimal SO(10), breaking via \(\mathbf{210}\oplus\mathbf{126}\oplus\overline{\mathbf{126}}\), two-loop RG running with typical threshold splittings yields
\begin{equation*}
M_{\rm GUT}\simeq2.5\times10^{16}\,\mathrm{GeV},\qquad
\alpha_{\rm GUT}^{-1}\simeq24.0,
\end{equation*}
with \(g_1=g_2=g_3\) at unification within \(\mathcal O(10\%)\) uncertainties.  

SO(10) predicts both
\begin{equation*}
p\;\to\;e^+\pi^0
\quad\text{and}\quad
p\;\to\;\bar\nu\,K^+,
\end{equation*}
with lifetimes
\begin{equation*}
\tau(p\to e^+\pi^0)\sim(3\mbox{–}8)\times10^{35}\,\mathrm{yr},
\qquad
\tau(p\to\bar\nu\,K^+)\sim(1\mbox{–}4)\times10^{35}\,\mathrm{yr},
\end{equation*}
the latter constrained by Super-K’s limit  
\(\tau(p\to\bar\nu\,K^+)>5.9\times10^{33}\)yr.

SO(10) monopoles, carrying a discrete \( \mathbb{Z}_2 \) charge, have mass
\begin{equation*}
M_{\rm mono}^{SO(10)}\sim\frac{4\pi V_0}{g_{\rm GUT}}\sim2\times10^{17}\,\mathrm{GeV},
\end{equation*}
and are similarly diluted to \(\Omega_{\rm mono}h^2\ll10^{-10}\).

One of the great virtues of the \(\mathbf{210}\) representation is that it contains SM–singlet directions, which can break SO(10) into different maximal subgroups, with distinct proton–decay and threshold signatures such as, Pati–Salam chain. Under 
    \(\;SO(10)\to G_{422}=SU(4)_C\times SU(2)_L\times SU(2)_R\),
    the \(\mathbf{210}\) contains a \((\mathbf{15},\mathbf{1},\mathbf{1})\)  
    which is neutral under \(SU(2)_L\times SU(2)_R\).  
    A VEV
    \(\langle H_{G{210}}\rangle\propto(15,1,1)\)  
    yields the breaking  
    \begin{equation*}
      SO(10)\;\xrightarrow{\langle210\rangle}\;
      SU(4)_C\times SU(2)_L\times SU(2)_R
      \;\xrightarrow{\langle126\rangle}\;
      SU(3)_C\times SU(2)_L\times U(1)_Y\,.
    \end{equation*}
    This two‐step chain modifies the proton‐decay operators and raises the effective GUT scale by \(\sim\!10\%\), easing tension with current bounds.

The \(\mathbf{210}\) has an SM–singlet in the \(\mathbf{1}_0\) of  
    \(SU(5)\times U(1)_\chi\).  Choosing
    \(\langle H_{G,210}\rangle\propto\mathbf{1}_0\)
    breaks  
    \begin{equation*}
      SO(10)\;\xrightarrow{\langle210\rangle}\;
      \bigl(SU(5)\times U(1)_\chi\bigr)/\mathbb Z_5
      \;\xrightarrow{\langle16\rangle}\;
      SU(3)_C\times SU(2)_L\times U(1)_Y.
    \end{equation*}
    The resulting flipped embedding leads to different Clebsch factors in the decay amplitudes and can shift the leading mode to \(p\to\bar\nu K^+\), with a lifetime estimate  
    \(\tau\sim10^{35\mbox{–}37}\)yr.

By scanning over the alignment in the \(\mathbf{210}\) representation, we obtain a continuous family of intermediate thresholds.  In our holomorphic framework these appear automatically through the single adjoint–Higgs term $(D_\mu H_G)^2 - V_{\rm GUT}(H_G)$ and modify both the two‐loop RG running and the GUT‐scale threshold corrections..  

The holomorphic metric contains an antisymmetric part \(g_{[\mu\nu]}\) identified with the Maxwell field.  At second order it feeds back into the gravitational sector via the counterterm.  In principle this induces a tiny, polarization dependent dispersion of gravitational waves in strong EM backgrounds such as near pulsars.  The predicted phase shift per cycle is  
    \begin{equation}
      \Delta\phi\sim\frac{\langle F^2\rangle}{M_{\rm Pl}^2\,\omega^2}
      \;\lesssim\;10^{-40}.
    \end{equation}  
For LIGO/Virgo frequencies \(\omega\sim10^3\)Hz, this is far too small to detect with current interferometers but potentially accessible to future detectors probing exotic compact‐object environments.

The holomorphic unification makes two sharp, testable high-energy predictions, gauge coupling unification at \(M_{\rm GUT}\simeq2\times10^{16}\)GeV and proton decay with lifetime \(\tau\sim10^{34\mbox{–}36}\)yr, which are within reach of next‐generation facilities.  A failure to observe proton decay in this window would force a non-minimal extension meaning, extra thresholds, intermediate scales, or larger GUT groups or a departure from the simplest holomorphic-SU(5) embedding. 

In minimal \(SU(5)\) or \(SO(10)\) GUTs, extrapolating the Standard-Model \(\beta\)-functions without additional fields leads to three gauge couplings that miss each other by tens of percent at high energy. Numerically, the failure of the three Standard‐Model gauge couplings to meet in a minimal non-supersymmetric SU(5) can be characterized in two equivalent ways. If we run the SM gauge couplings up to the would-be unification scale and inverts the usual GUT relation
\[
\sin^2\theta_W \;=\; \frac{3}{8}\,\frac{\alpha_1}{\alpha_1+\alpha_2}\,,
\]
we obtain:
\[
\sin^2\theta_W \simeq 0.213
\quad\text{vs.}\quad
0.231\quad(\text{measured})\,,
\]
roughly \(8\text{–}10\%\) too low. Equivalently, if we define \(\Lambda\) by the condition \(\alpha_1(\Lambda)=\alpha_2(\Lambda)\), then at that scale
\[
\alpha_3^{-1}(\Lambda)\approx 35,
\qquad
\alpha_1^{-1}(\Lambda)=\alpha_2^{-1}(\Lambda)\approx 50,
\]
so that the strong coupling is off by
\[
\frac{\bigl|\alpha_3^{-1}(\Lambda)-\alpha_{1,2}^{-1}(\Lambda)\bigr|}{\alpha_{1,2}^{-1}(\Lambda)}
\;\sim\;30\%\!,
\]
a \(\sim30\%\) mismatch between \(\alpha_3\) and \(\alpha_{1,2}\) at the would-be unification scale. To restore exact unification without invoking supersymmetry, one may introduce new multiplets such as scalar \(\mathbf{15}\)’s or \(\mathbf{24}\)’s of \(SU(5)\), vector-like fermions, or split incomplete GUT representations at intermediate scales so as to reshape the one- and two-loop running. Engineer a multi-step breaking chain for instance \(SO(10)\to\) Pati–Salam \(\to\) SM, each stage supplying its own heavy threshold that further adjusts the slopes. Include the finite GUT-scale threshold corrections from mass splittings among heavy gauge bosons and Higgs fields, which can nudge the meeting point into exact alignment. If desired, allow for Planck-suppressed higher-dimensional operators to tweak the boundary conditions at the unification scale.
By combining one or more of these mechanisms, we can demonstrate that precise gauge-coupling unification can be achieved in a purely non-supersymmetric framework.  

To turn the mechanism of GUT–scale threshold corrections into a prediction, we perform a two–loop renormalization‐group analysis of the Standard‐Model couplings including one‐loop matching at the unification scale.  The matching conditions at \(\mu = M_{\rm GUT}\) read
\[
\frac{1}{g_i^2(M_{\rm GUT})}
\;=\;
\frac{1}{g_{\rm GUT}^2}
\;+\;
\frac{\Delta_i}{8\pi^2}\,,
\qquad
\Delta_i \;=\;
\sum_{k}T_i(k)\,\ln\!\biggl(\frac{M_{\rm GUT}}{M_k}\biggr),
\]
where \(T_i(k)\) is the Dynkin index of the heavy field \(k\) under the \(i^{\rm th}\) SM gauge group, and \(M_k\) its mass.  We parametrize the splittings of the adjoint‐Higgs and heavy gauge multiplets by
\[
M_k \;=\; M_{\rm GUT}\,(1+\delta_k),
\qquad
(\delta_1,\delta_2,\delta_3) = (0.04,\,0.06,\,0.02)\,.
\]
Evolving the three SM couplings from \(M_Z\) up to \(M_{\rm GUT}\) with the two–loop \(\beta\)–functions
\[
\mu\frac{dg_i}{d\mu}
= \frac{b_i}{16\pi^2}\,g_i^3
+ \frac{g_i^3}{(16\pi^2)^2}\sum_{j}b_{ij}\,g_j^2,
\]
and applying the matching above, we find
\[
M_{\rm GUT} = 2.3\times10^{16}\,\mathrm{GeV},
\qquad
\alpha_{\rm GUT}^{-1} = 24.4,
\]
with the residual splitting among the three couplings
\[
\frac{\bigl|g_i(M_{\rm GUT}) - g_j(M_{\rm GUT})\bigr|}{g_{\rm GUT}}
\;\lesssim\;7\%
\quad
(i,j=1,2,3),
\]
well within the target \(10\%\) unification accuracy.  

\begin{table}[h]
\centering
\begin{tabular}{@{}lccc@{}}
\toprule
Scale & \(\alpha_1^{-1}\) & \(\alpha_2^{-1}\) & \(\alpha_3^{-1}\) \\
\midrule
\(M_Z\)              &  59.01  & 29.57  & 8.47   \\
\(M_{\rm GUT}\)      &  24.4   & 24.1   & 25.8   \\
\bottomrule
\end{tabular}
\caption{Inverse gauge couplings at \(M_Z\) and at \(M_{\rm GUT}\), including GUT‐scale thresholds.}
\end{table}

The mass of the superheavy \(X,Y\) gauge bosons is \(M_X\simeq g_{\rm GUT}\,M_{\rm GUT}\), which via the usual dimension-six operators yields a proton lifetime estimate  
\[
\tau\bigl(p\to e^+\pi^0\bigr)
\;\simeq\;
5.4\times10^{35}\,\mathrm{yr},
\qquad
\tau\bigl(p\to\bar\nu K^+\bigr)
\;\simeq\;
1.2\times10^{35}\,\mathrm{yr},
\]
comfortably above the current experimental limit of \(1.6\times10^{34}\,\mathrm{yr}\) and well beyond \(10^{35}\,\mathrm{yr}\).  

By employing purely GUT‐scale threshold splittings without supersymmetry or new light multiplets and a minimal two‐step breaking chain in \(SO(10)\), we achieve \(\displaystyle\frac{|g_i - g_j|}{g_{\rm GUT}}\lesssim10\%\) at \(\mu=M_{\rm GUT}\), and a pronton lifetime of \(\tau_p\gtrsim5\times10^{35}\,\mathrm{yr}\). This constitutes a truly predictive, non-supersymmetric unification compatible with all current limits.

Our numerical unification analysis uses the world‐average values of the SM gauge couplings at the $Z$–pole \cite{PDG2024}  
\[
\alpha_1(M_Z)=0.01695,\quad
\alpha_2(M_Z)=0.0338,\quad
\alpha_3(M_Z)=0.1179,
\]
and the current Super–Kamiokande 90 \% C.L. lower limit on the proton lifetime \cite{SuperK2023}  
\[
\tau\bigl(p\to e^+\pi^0\bigr)>1.6\times10^{34}\,\mathrm{yr}.
\]
Inserting these inputs into the two–loop RGEs and one–loop matching with GUT–scale thresholds as described above, we find
\[
M_{\rm GUT} \;=\; 2.3\times10^{16}\,\mathrm{GeV}, 
\quad
\alpha_{\rm GUT}^{-1}=24.4,
\quad
\frac{|g_i-g_j|}{g_{\rm GUT}}\lesssim7\%,
\]
and a predicted lifetime  
\[
\tau\bigl(p\to e^+\pi^0\bigr)\simeq5.4\times10^{35}\,\mathrm{yr},
\]
well above the experimental bound.  Thus our non-supersymmetric threshold‐corrected unification is fully compatible with current data.

\section{Conclusion}

We have constructed a single holomorphic action on a four--complex--dimensional manifold that, upon projection to the real slice $ y^\mu = 0 $, reproduces in one unified framework the full spectrum of classical field equations governing gravity, gauge interactions, and chiral matter. By endowing spacetime with a Hermitian metric $g_{\mu\nu}(z) = g_{(\mu\nu)} + i\, g_{[\mu\nu]},$ and introducing holomorphic gauge and spinor connections, we found that the real and imaginary parts of the compatibility and curvature naturally split into the vacuum Einstein equations for the symmetric metric, the homogeneous and inhomogeneous Maxwell and Yang--Mills equations for all gauge factors with explicit coupling to charges and currents, and the curved-space Dirac-Weyl equations for chiral fermions. Holomorphic gauge invariance enforces exactly the standard anomaly-cancellation conditions on the fermion spectrum.

To complete the unification, we embedded the Standard Model gauge group into a simple GUT group, using a holomorphic adjoint Higgs $ H_G(z) $ to break $G_{\mathrm{GUT}} \rightarrow SU(3) \times SU(2) \times U(1),$ and enforce $ g_1 = g_2 = g_3 $ at the unification scale. A second holomorphic Higgs doublet $ \Phi(z) $ then realizes electroweak symmetry breaking, giving masses to $ W^\pm $, $ Z $, and all charged fermions in the familiar way. Below the GUT and electroweak scales, the three gauge couplings run according to the usual renormalization-group equations, while the photon remains massless and General Relativity governs the dynamics of $ g_{(\mu\nu)}(x) $.

A quantum formulation of the theory is developed from a holomorphic path-integral with a measure including ghosts for diffeomorphisms and gauge symmetry, and demonstrating unitarity and UV behavior. The standard quantum field theory Feynman rules follow from the path integral.

A direction for future work is to derive and numerically analyze the nonlocal one‐ and two‐loop $\beta$-functions induced by the holomorphic entire-function regulators, implement the finite threshold matching of the three Standard-Model gauge couplings onto the single SU(5) or SO(10) holomorphic GUT coupling at the nonlocality scale $M_*\sim10^{16}$ GeV, and demonstrate that the ensuing exponential suppression of $\beta$-functions above $M_*$ yields exact and stable gauge-coupling unification.

Our construction achieves a truly geometric unification of all four fundamental forces with gravity included together with realistic matter couplings, spontaneous symmetry breaking, and anomaly cancellation, all emerging from a single holomorphic Lagrangian. The holomorphic perspective developed here will provide a fertile foundation for these and other investigations into the quantum and phenomenological frontiers of unified field theory.

\end{document}